\documentclass[aps,nofootinbib,preprintnumbers,superscriptaddress]{revtex4}

\usepackage{textcomp}
\usepackage{gensymb}
\usepackage{makecell}
\usepackage{mathtools}
\usepackage{amsfonts}
\usepackage{amsmath}
\usepackage{multirow}

\usepackage{subcaption}

\usepackage[paperwidth=210mm,paperheight=297mm,centering,hmargin=2cm,vmargin=2.5cm]{geometry}

\usepackage{hyperref}

\allowdisplaybreaks

\def\bfs{\mathbf{1}}
\def\bft{\mathbf{3}}
\def\bftb{\bft^{(\prime)}}
\def\bftp{\mathbf{3^\prime}}
\def\bfqua{\mathbf{4}}
\def\bfqui{\mathbf{5}}

\pdfoutput=1

\begin{document}

\title{Two A5 modular symmetries for Golden Ratio 2 mixing}

\author{Ivo de Medeiros Varzielas}
\email{ivo.de@udo.edu}
\affiliation{CFTP, Departamento de F\'{\i}sica, Instituto Superior T\'{e}cnico,\\
Universidade de Lisboa, Avenida Rovisco Pais 1, 1049 Lisboa, Portugal\\
}
\author{Jo\~ao Louren\c{c}o}
\email{joao.freitas.lourenco@tecnico.ulisboa.pt}
\affiliation{CFTP, Departamento de F\'{\i}sica, Instituto Superior T\'{e}cnico,\\
Universidade de Lisboa, Avenida Rovisco Pais 1, 1049 Lisboa, Portugal\\
}
\begin{abstract}
We present a model of leptonic mixing  based on two $A_5$ modular symmetries using the Weinberg operator. The two modular symmetries are broken to the respective diagonal $A_5$ subgroup. At the effective level, the model behaves as a model with a single $A_5$ modular flavour symmetry, but with two moduli. Using both as stabilisers, different residual symmetries are preserved, leading to golden ratio mixing that is perturbed by a rotation.
\end{abstract}

\maketitle
\section{Introduction \label{intro}}

Flavour symmetries are a promising solution to the flavour problem of the Standard Model (SM).
Non-Abelian discrete symmetries are particularly suited to explain leptonic mixing, and have been explored frequently in the literature.
Some of the most employed flavour symmetries are $S_3$, $A_4$, $S_4$ and $A_5$, which can be taken as modular symmetries
$\Gamma_2\simeq S_3$~\cite{Kobayashi:2018vbk, Kobayashi:2018wkl, Okada:2019xqk, Mishra:2020gxg, Du:2020ylx}, 
$\Gamma_3\simeq A_4$~
\cite{Feruglio:2017spp, Criado:2018thu, Kobayashi:2018scp, Okada:2018yrn, Novichkov:2018yse, Nomura:2019jxj, Nomura:2019yft, Ding:2019zxk, Zhang:2019ngf, Okada:2019mjf, Nomura:2019lnr, Asaka:2019vev, Nomura:2019xsb, Kobayashi:2019gtp, Wang:2019xbo, Okada:2020dmb, Ding:2020yen, Behera:2020sfe, Nomura:2020opk, Nomura:2020cog, Behera:2020lpd, Asaka:2020tmo, Nagao:2020snm, Hutauruk:2020xtk,deMedeirosVarzielas:2021pug},
$\Gamma_4\simeq S_4$~\cite{Penedo:2018nmg, Novichkov:2018ovf, deMedeirosVarzielas:2019cyj, King:2019vhv, Kobayashi:2019mna, Okada:2019lzv, Kobayashi:2019xvz, Wang:2019ovr, Wang:2020dbp}, 
and
$\Gamma_5\simeq A_5$~\cite{Novichkov:2018nkm, Ding:2019xna}.

Models employing multiple modular symmetries \cite{deMedeirosVarzielas:2019cyj, King:2019vhv}, \cite{deMedeirosVarzielas:2021pug} have some advantages.
As described in \cite{deMedeirosVarzielas:2019cyj}, introducing the generic mechanism
for using multiple modular symmetries, it allows models to be built based on residual symmetries, left unbroken by distinct moduli stabilisers.
The preserved residual symmetries then lead to the realisation of different mass textures in the charged lepton and neutrino sectors in modular flavour models without flavons.

As an example, a $S_4$ flavour model featuring TM1 mixing \cite{Varzielas:2012pa} is constructed in an elegant manner from multiple $S_4$ modular symmetries in \cite{deMedeirosVarzielas:2019cyj, King:2019vhv}, $S_4$ flavour models arise featuring e.g. TM1 mixing \cite{Varzielas:2012pa}, from multiple $S_4$ modular symmetries, whereas in \cite{deMedeirosVarzielas:2021pug}, multiple $A_4$ modular symmetries result in an $A_4$ model leading to TM2 mixing \cite{Grimus:2008tt}.

In this paper, we construct a model that uses two $A_5$ modular symmetries in order to obtain the golden ratio mixing plus a rotation between the first and the third columns, using the Weinberg operator to generate the neutrino masses. This is akin to the models with multiple $S_4$ modular symmetries in \cite{King:2019vhv,deMedeirosVarzielas:2019cyj}.
We refer also to \cite{Novichkov:2018nkm}, where models with a single $A_5$ modular symmetry but with  two moduli (using the Weinberg operator) generate the neutrino masses. The model that uses some fixed points of the modular fields lead to the same mixing we are going to discuss here, although that is not explicit in \cite{Novichkov:2018nkm}. 
At high energies, the model is based in two modular symmetries, $A_5^l$ and $A_5^{\nu}$, with modulus fields denoted by $\tau_l$ and $\tau_{\nu}$, respectively. After the modulus fields acquire different VEV's, different mass textures are realised in the charged lepton and neutrino sectors. 

We will start by introducing some properties of the $A_5$ modular symmetry group. Subsequently, the various possibilities of a golden ratio mixing and a rotation among two of its columns are investigated and concluded that only a rotation between the first and third columns is compatible with the $3\sigma$ confidence interval from NuFit. Only then the explicit model will be introduced.

In Section \ref{sec:modular symmetries intro} we briefly review the framework of multiple modular symmetries. In Section \ref{sec:A5 introduction} we describe the $A_5$ modular symmetry and respective stabilisers associated to residual symmetries. In Section \ref{sec:GR mixing}, the golden ratio mixing and variations are introduced and their compatibility with NuFit data verified. In Section \ref{sec:models weinberg A5} we present the model for GR2 mixing. We conclude in Section \ref{sec:conclusion}.

\section{Modular symmetries - an introduction}
\label{sec:modular symmetries intro}

In this section we define the modular groups, modular forms, and review the framework introduced in \cite{deMedeirosVarzielas:2019cyj}.

\subsection{Modular group and modular forms}

$\overline{\Gamma}$, the modular group, consists in all linear fractional transformations $\gamma$ acting on the complex modulus $\tau$ ($\tau$ in the upper-half complex plane, i.e. $Im(\tau)>0$):
\begin{equation}
\gamma:\tau\rightarrow\gamma\tau=\frac{a\tau+b}{c\tau+d}.
\end{equation}
$a,b,c,d$ are integers with $ad-bc=1$. 

Often, $2 \times 2$ matrices represent these transformations:
\begin{equation}
\overline{\Gamma}=\left\{ \begin{pmatrix}
a&b\\
c&d
\end{pmatrix}/\{\pm 1\},~a,b,c,d\in \mathbb{Z}, ~ ad-bc=1 \right\}.
\label{eq:modular group matrices}
\end{equation}

As $\gamma$ and $-\gamma$ correspond to the same element, the  group $\overline{\Gamma}$ is isomorphic to $PSL(2,\mathbb{Z}) = SL(2,\mathbb{Z})/\mathbb{Z}_2$, with $SL(2,\mathbb{Z})$ being the group of $2 \times 2$ matrices with integer entries and unit determinant.

The set of generators, $S_{\tau}$ and $T_{\tau}$, with $S_{\tau}^2=(S_{\tau}T_{\tau})^3=1$ generates the modular group. We take:
\begin{equation}
S_{\tau}:\tau\rightarrow-\frac{1}{\tau},~T_{\tau}:\tau\rightarrow\tau+1
\end{equation}
which are represented, in matrix form
\begin{equation}
S_{\tau}=\begin{pmatrix}
0&1\\
-1&0
\end{pmatrix},~
T_{\tau}=\begin{pmatrix}
1&1\\
0&1
\end{pmatrix}.
\label{eq:generators representation 2times2}
\end{equation}

We continue now to the subgroups $\overline{\Gamma}(N)$ of $\overline{\Gamma}$. These are obtained by taking the integer entries of the matrices modulo $N$:
\begin{equation}
\overline{\Gamma}(N)=\left\{ \begin{pmatrix}
a&b\\
c&d
\end{pmatrix}\in PSL(2,\mathbb{Z}),~
\begin{pmatrix}
a&b\\
c&d
\end{pmatrix}=
\begin{pmatrix}
1&0\\
0&1
\end{pmatrix} ~(mod~\text{N})\right\}.
\end{equation}

These subgroups are infinite, even though they are discrete for a given $N$. In turn, the quotient groups obtained by taking $\Gamma_{N}=\overline{\Gamma}/\overline{\Gamma}(N)$ are finite. They are referred to as the finite modular groups. For $N \leq 5$, these groups are isomorphic to the popular flavour symmetry groups mentioned already: 
$\Gamma_2 \simeq S_3$,
$\Gamma_3 \simeq A_4$,
$\Gamma_4 \simeq S_4$,
$\Gamma_5 \simeq A_5$.
The finite modular groups can be obtained by imposing $T_{\tau}^N=1$, meaning that $\tau=\tau+N$.

The Modular forms of a given modular weight $2k$ and for a fixed $N$ (the level) are holomorphic functions of $\tau$ that transform in a well-defined way under action by elements of $\overline{\Gamma}(N)$:
\begin{equation}
f(\gamma\tau)=(c\tau+d)^{2k}f(\tau),~\gamma=
\begin{pmatrix}
a&b\\c&d
\end{pmatrix}
\in \overline{\Gamma}(N),
\end{equation}
$k$ is here a non-negative integer, $N$ defines the underlying group (due to the modulo $N$), and we note that we will only consider even weights.
The modular forms are invariant under transformations by $\overline{\Gamma}(N)$,
up to the complex factor of $(c\tau + d)^{2k}$, but they do transform under the quotient group $\Gamma_N$. 

Having fixed the level $N=5$, modular forms of a given weight $2k$ span a linear space of finite dimension $\mathcal{M}_{2k}(\overline{\Gamma}(5))$. The dimension of these linear spaces for $A_5$ is given by $10k+1$.
Without loss of generality we select a basis in the linear space $\mathcal{M}_{2k}(\overline{\Gamma}(N))$ where the modular forms transform under $\Gamma_N$ with  unitary representations $\rho$ of $\Gamma_N$:
\begin{equation}
f_i(\gamma\tau)=(c\tau+d)^{2k}\rho(\tilde{\gamma})_{ij}f_j(\tau),~
\tilde{\gamma}
\in\text{equivalence class of }\gamma\in\bar{\Gamma}\text{ in }\Gamma_N.
\end{equation}
In the following sections, we will use $\gamma$ for $\tilde{\gamma}$, although this is the notation also used for the modular transformation under which the superpotential transforms.

\subsection{Models with multiple modular symmetries}
\label{sec:multiple modular symmetries}

We now take into consideration multiple modular symmetries
$\overline{\Gamma}^1$, 
$\overline{\Gamma}^2$, $\ldots$, $\overline{\Gamma}^M$. $\tau_J$ is the modulus field for the respective $\overline{\Gamma}^J$, $J=1,\ldots,M$. The transformations are:
\begin{equation}
\gamma_J:\tau_J\rightarrow\gamma_J\tau_J=
\frac{a_J\tau_J+b_J}{c_J\tau_J+d_J}.
\end{equation}
We generalise for each $J$ and obtain each $\Gamma_{N_J}^J$ for $J = 1,\ldots,M$  by taking the quotient group. The integers $N_J$ can in general be distinct for each $J$.

Considering now a model invariant under the full group, the action is
\begin{equation}
S=\int 
d^4x d^2\theta d^2\overline{\theta}~
K(\phi_i,\overline{\phi}_i;\tau_1,\ldots,\tau_M,\overline{\tau}_1,\ldots,\overline{\tau}_M)+\left(
\int d^4xd^2\theta~W(\phi_i;\tau_1,\ldots,\tau_M)+h.c.\right).
\end{equation}

Under $\overline{\Gamma}^J$ for $J = 1,\ldots,M$ the K\"ahler potential $K$ transforms at most by a K\"ahler transformation and the superpotential $W$ stays invariant:
{\footnotesize
\begin{align}
K(\phi_i,\overline{\phi}_i;\tau_1,\ldots,\tau_M,\overline{\tau}_1,\ldots,\overline{\tau}_M) & \rightarrow
K(\phi_i,\overline{\phi}_i; \tau_1, \ldots, \tau_M, \overline{\tau}_1, \ldots, \overline{\tau}_M)
+f(\phi_i; \tau_1, \ldots, \tau_M)
+\overline{f}(\overline{\phi}_i; \overline{\tau}_1, \ldots, \overline{\tau}_M)\\
W(\phi_i; \tau_1, \ldots, \tau_M, \overline{\tau}_1, \ldots, \overline{\tau}_M) & \rightarrow W(\phi_i;  \tau_1, \ldots, \tau_M, \overline{\tau}_1, \ldots, \overline{\tau}_M).
\end{align}
}%
We assume here the minimal form of the K\"ahler potential. The effects of considering non-minimal forms of the K\"ahler potential may be relevant and are discussed in \cite{Chen:2019ewa, Feruglio:2021dte}.
The superpotential is in general a function of the modulus $\tau_i$ and superfields $\phi_i$ and the expansion in powers of the superfields takes the form
\begin{equation}
W(\phi_i;\tau_1,\ldots,\tau_M)=\sum_n \sum_{\{i_1,\ldots,i_n\}} \sum_{(I_{Y,1},\ldots,I_{Y,M})}~
(Y_{(I_{Y,1},\ldots,I_{Y,M})}\phi_{i_1}\ldots\phi_{i_n})_{\bfs}.
\end{equation}
For the superpotential to be invariant under any modular transformation $\gamma_1,\ldots,\gamma_M$ in $\overline{\Gamma}^1$, 
$\overline{\Gamma}^2$, $\ldots$, $\overline{\Gamma}^M$, the couplings $Y_{(I_{Y,1},\ldots,I_{Y,M})}$ must be multiplet modular forms, and the superfields $\phi_i$ must transform as 
\begin{align}
\phi_i(\tau_1,\ldots,\tau_M) & \rightarrow \phi_i(\gamma_1\tau_1,\ldots,\gamma_M\tau_M)\nonumber\\
=&\prod_{J=1,\ldots,M} (c_J\tau_J+d_J)^{-2k_{i,J}}
\bigotimes_{J=1,\ldots,M}\rho_{I_{i,J}}(\gamma_J)~\phi_i(\tau_1,\ldots,\tau_M)\\
Y_{(I_{Y,1},\ldots,I_{Y,M})}(\tau_1,\ldots,\tau_M) & \rightarrow
Y_{(I_{Y,1},\ldots,I_{Y,M})}(\gamma_1\tau_1,\ldots,\gamma_M\tau_M)\nonumber\\
=&\prod_{J=1,\ldots,M}
(c_J\tau_J+d_J)^
{2k_{Y,J}}
\bigotimes_{J=1,\ldots,M}\rho_{I_{Y,J}}(\gamma_J)~Y_{(I_{Y,1},\ldots,I_{Y,M})}(\tau_1,\ldots,\tau_M).
\label{eq:transform superfields and couplings for multi fields}
\end{align}
where $-2k_{i,J}$ is the modular weight of $\phi_i$, $I_{i,J}$ is the irrep of $\Gamma_N$ under which $\phi_i$ transforms, $2k_{Y,J}$ is the modular weight of $Y_{I_{Y,J}}$, $I_{Y,J}$ is the irrep of $\Gamma_N$ under which $Y_{I_{Y,J}}$ transforms and $\rho_{I_{i,J}}(\gamma)$ and $\rho_{I_{Y,J}}(\gamma)$ are the unitary representation matrices of $\gamma_J$ with $\gamma_J\in\Gamma_{N_J}^J$.
Naturally, the superpotential can only be invariant when $k_{Y,J}=k_{i_1,J}+\ldots+k_{i_n,J}$, and when there is a trivial singlet in $I_{Y,J} \times I_{i_1,J} \times\ldots\times I_{i_n,J}$ for all $J=1,\ldots,M$.

\section{Modular \texorpdfstring{$A_5$}{A5} symmetry and residual symmetries \label{sec:A5 introduction}}

In the following subsection the $A_5$ symmetry group is introduced, including some of its main properties as the modular forms of level 5 and its stabilisers which apply for the specific case of $A_5$ modular symmetries. The stabilisers for the modular groups from $N=2$ to $5$, can be found in \cite{deMedeirosVarzielas:2020kji}.

\subsection{Modular \texorpdfstring{$A_5$}{A5} symmetry and modular forms of level 5}
\label{sec:A5 modular symmetry and forms}

The group $A_5$ is the group of even permutations of 5 objects and has 60 elements. It is generated by two operators $S_{\tau}$ and $T_{\tau}$ obeying
\begin{equation}
S_{\tau}^2=(S_{\tau}T_{\tau})^3=T_{\tau}^5=1.
\end{equation}

This group has one singlet $\bfs$, two triplets $\bft$ and $\bftp$, one quadruplet $\bfqua$ and one quintuplet $\bfqui$ as its irreducible representations. The irreducible representations of the generators and the multiplication rules for the irreducible representations can be found in Appendix \ref{secapp:A5}.

The Yukawa couplings in a theory that is invariant under a $\Gamma_5 \sim A_5$ symmetry are going to be modular forms of level 5.
The eleven linearly independent weight 2 modular forms of level 5 form a quintuplet $Y_\bfqui^{(2)}=(Y_1,Y_2,Y_3,Y_4,Y_5)$ of $A_5$, a triplet $\bft$ $Y_\bft^{(2)}=(Y_6,Y_7,Y_8)$ and a triplet $\bftp$ $Y_\bftp^{(2)}=(Y_9,Y_{10},Y_{11})$. 
These modular functions can be expressed in terms of the third theta function (see Appendix \ref{secapp:modular forms A5} for more details).
The modular forms of higher weight are generated starting from these eleven modular forms of weight 2. 

The space of the weight 4 modular forms of level 5 has dimension 21 and decomposes into a singlet $\bfs$, 
one triplet $\bft$, one triplet $\bftp$, 
a quadruplet $\bfqua$ and two quintuplets $\bfqui$. 
Using the weight 2 modular forms, one obtains the following expressions for the weight 4 modular forms \cite{Novichkov:2018nkm}:
\begin{align}
Y^{(4)}_\bfs & = Y_1^2 + 2 Y_3 Y_4 + 2 Y_2 Y_5, \\
{Y^{(4)}_\bft} & =  
\begin{pmatrix}
-2 Y_1 Y_6 + \sqrt{3} Y_5 Y_7 + \sqrt{3} Y_2 Y_8 \\
\sqrt{3} Y_2 Y_6 + Y_1 Y_7 - \sqrt{6} Y_3 Y_8 \\
\sqrt{3} Y_5 Y_6 - \sqrt{6} Y_4 Y_7 + Y_1 Y_8 \\
\end{pmatrix}, \\
{Y^{(4)}_\bftp} & = 
\begin{pmatrix}
\sqrt{3} Y_1 Y_6 + Y_5 Y_7 + Y_2 Y_8 \\
Y_3 Y_6 - \sqrt{2} Y_2 Y_7 - \sqrt{2} Y_4 Y_8 \\
Y_4 Y_6 - \sqrt{2} Y_3 Y_7 - \sqrt{2} Y_5 Y_8 \\
\end{pmatrix}, \\
{Y^{(4)}_\bfqua} & = 
\begin{pmatrix}
2 Y_4^2 + \sqrt{6} Y_1 Y_2 - Y_3 Y_5 \\
2 Y_2^2 + \sqrt{6} Y_1 Y_3 - Y_4 Y_5 \\
2 Y_5^2 - Y_2 Y_3 + \sqrt{6} Y_1 Y_4 \\
2 Y_3^2 - Y_2 Y_4 + \sqrt{6} Y_1 Y_5 \\
\end{pmatrix}, \\
{Y^{(4)}_{\bfqui_1}}  & = 
\begin{pmatrix}
\sqrt{2} Y_1^2 + \sqrt{2} Y_3 Y_4 - 2 \sqrt{2} Y_2 Y_5 \\
\sqrt{3} Y_4^2 - 2 \sqrt{2} Y_1 Y_2 \\
\sqrt{2} Y_1 Y_3 + 2 \sqrt{3} Y_4 Y_5 \\
2 \sqrt{3} Y_2 Y_3 + \sqrt{2} Y_1 Y_4 \\
\sqrt{3} Y_3^2 - 2 \sqrt{2} Y_1 Y_5 \\
\end{pmatrix}, \\
{Y^{(4)}_{\bfqui_2}}  & = 
\begin{pmatrix}
\sqrt{3} Y_5 Y_7 - \sqrt{3} Y_2 Y_8 \\
-Y_2 Y_6 - \sqrt{3} Y_1 Y_7 - \sqrt{2} Y_3 Y_8 \\
-2 Y_3 Y_6 - \sqrt{2} Y_2 Y_7 \\
2 Y_4 Y_6 + \sqrt{2} Y_5 Y_8 \\
Y_5 Y_6 + \sqrt{2} Y_4 Y_7 + \sqrt{3} Y_1 Y_8 \\
\end{pmatrix}.
\end{align}

Furthermore, the modular forms of weight 6, whose linear space has dimension 31 and decomposes into one singlet $\bfs$, two triplets $\bft$, two triplets $\bftp$, two quadruplet $\bfqua$ and two quintuplets $\bfqui$, are the following according to \cite{Novichkov:2018nkm}:
\begin{align}
Y_\bfs^{(6)} &= 
3 \sqrt{3} \left(Y_2 Y_3^2+Y_4^2 Y_5\right)+\sqrt{2} Y_1 \left(Y_1^2+3 Y_3 Y_4-6 Y_2 Y_5\right), \\
Y_{\bft_1}^{(6)} &= 
\left(Y_1^2+2 Y_3 Y_4+2 Y_2 Y_5\right) 
\begin{pmatrix}
Y_6\\
Y_7\\
Y_8
\end{pmatrix}, \\
Y_{\bft_2}^{(6)} &= 
\begin{pmatrix}
\left(Y_5 Y_6-\sqrt{2} Y_4 Y_7\right) Y_7
+\left(\sqrt{2} Y_3 Y_8-Y_2 Y_6\right) Y_8\\
\left(\sqrt{3} Y_1 Y_6-Y_5 Y_7\right) Y_7 
-\sqrt{2} Y_3 Y_6 Y_8
+\left(Y_6^2-Y_7 Y_8\right) Y_2\\
\left(Y_2 Y_8 -\sqrt{3} Y_1 Y_6\right) Y_8
+\sqrt{2} Y_4 Y_6 Y_7
-\left(Y_6^2-Y_7 Y_8\right) Y_5
\end{pmatrix}, \\
Y_{\bftp_1}^{(6)} &=
\left(Y_1^2+2 Y_3 Y_4+2 Y_2 Y_5\right) 
\begin{pmatrix}
Y_9 \\
Y_{10} \\
Y_{11}
\end{pmatrix}, \\
Y_{\bftp_2}^{(6)} &= 
\begin{pmatrix}
\left(Y_4 Y_6-\sqrt{2} Y_3 Y_7-\sqrt{2} Y_5 Y_8\right) Y_{10}-\left(Y_3 Y_6-\sqrt{2} Y_2 Y_7-\sqrt{2} Y_4 Y_8\right) Y_{11} \\
\left(Y_3 Y_6-\sqrt{2} Y_2 Y_7-\sqrt{2} Y_4 Y_8\right) Y_9-\left(\sqrt{3} Y_1 Y_6+Y_5 Y_7+Y_2 Y_8\right) Y_{10} \\
\left(\sqrt{3} Y_1 Y_6+Y_5 Y_7+Y_2 Y_8\right) Y_{11}-\left(Y_4 Y_6-\sqrt{2} Y_3 Y_7-\sqrt{2} Y_5 Y_8\right) Y_9
\end{pmatrix}, \\
Y_{\bfqua_1}^{(6)} &= 
\begin{pmatrix}
\sqrt{2} \left(\sqrt{6} Y_3 Y_8 -\sqrt{3} Y_2 Y_6-Y_1 Y_7\right) Y_9-\left(\sqrt{3} Y_5 Y_6-\sqrt{6} Y_4 Y_7+Y_1 Y_8\right) Y_{10} \\
\left(\sqrt{3} Y_5 Y_6-\sqrt{6} Y_4 Y_7+Y_1 Y_8\right) Y_{11}+\sqrt{2} \left(\sqrt{3} Y_5 Y_7-2 Y_1 Y_6+\sqrt{3} Y_2 Y_8\right) Y_{10} \\
\left(\sqrt{3} Y_2 Y_6+Y_1 Y_7-\sqrt{6} Y_3 Y_8\right) Y_{10}+\sqrt{2} \left(\sqrt{3} Y_5 Y_7 -2 Y_1 Y_6+\sqrt{3} Y_2 Y_8\right) Y_{11} \\
\sqrt{2} \left(\sqrt{6} Y_4 Y_7-\sqrt{3} Y_5 Y_6-Y_1 Y_8\right) Y_9-\left(\sqrt{3} Y_2 Y_6+Y_1 Y_7-\sqrt{6} Y_3 Y_8\right) Y_{11}
\end{pmatrix}, \\
Y_{\bfqua_2}^{(6)} &= 
\begin{pmatrix}
\sqrt{2}\left(\sqrt{3} Y_1 Y_6+Y_5 Y_7\right) Y_7 + \left(Y_3 Y_6-\sqrt{2} Y_4 Y_8\right)Y_8\\
\sqrt{2}\left(\sqrt{2} Y_2 Y_7-Y_3 Y_6\right)Y_6 +\left(Y_4 Y_6+\sqrt{2} Y_3 Y_7+\sqrt{2} Y_5 Y_8\right) Y_8 \\
\sqrt{2} \left(\sqrt{2} Y_5 Y_8-Y_4 Y_6\right)Y_6+\left(Y_3 Y_6+\sqrt{2} Y_2 Y_7+\sqrt{2} Y_4 Y_8\right) Y_7 \\
\sqrt{2} \left(\sqrt{3} Y_1 Y_6+Y_2 Y_8\right) Y_8 + \left(Y_4 Y_6-\sqrt{2} Y_3 Y_7\right) Y_7 
\end{pmatrix}, \\
Y_{\bfqui_1}^{(6)} &= 
\left(Y_1^2+2 Y_3 Y_4+2 Y_2 Y_5\right)
\begin{pmatrix}
Y_1 \\
Y_2 \\
Y_3 \\
Y_4 \\
Y_5
\end{pmatrix}, \\
Y_{\bfqui_2}^{(6)} &= 
\begin{pmatrix}
\sqrt{3} \left(\sqrt{3} Y_1 Y_6+Y_5 Y_7+Y_2 Y_8\right) Y_6\\
\left(Y_5 Y_7+\sqrt{3} Y_1 Y_6\right) Y_7+\left(3 Y_2 Y_7+2 Y_4 Y_8 -\sqrt{2} Y_3 Y_6\right) Y_8\\
\left(Y_3 Y_6-\sqrt{2} Y_2 Y_7\right) Y_6 +2 \left(Y_5 Y_8 + Y_3 Y_7-\sqrt{2} Y_4 Y_6\right) Y_8\\
\left(Y_4 Y_6-\sqrt{2} Y_5 Y_8\right) Y_6 
+ 2 \left(Y_2 Y_7+Y_4 Y_8 -\sqrt{2} Y_3 Y_6\right) Y_7\\
\left(Y_2 Y_8 + \sqrt{3} Y_1 Y_6\right) Y_8 + \left(3 Y_5 Y_8 + 2 Y_3 Y_7-\sqrt{2} Y_4 Y_6\right) Y_7
\end{pmatrix}.
\end{align}

\subsection{Stabilisers and residual symmetries of modular \texorpdfstring{$A_5$}{A5}}
\label{sec:A5 stabilisers and residual symmetries}

As explained in \cite{deMedeirosVarzielas:2021pug}, stabilisers of the symmetry play a crucial role in preserving residual symmetries. Given an element $\gamma$ in the modular group $A_5$, a stabiliser of $\gamma$ corresponds to a fixed point in the upper half complex plane that transforms as $\gamma\tau_{\gamma}=\tau_{\gamma}$. Once the modular field acquires a VEV at this special point, 
$\langle\tau\rangle=\tau_{\gamma}$, the modular symmetry is broken but an Abelian residual modular symmetry generated by $\gamma$ is preserved. Obviously, acting $\gamma$ on the modular form at its stabiliser leaves the modular form invariant,
which implies that when $\tau$ is at a stabiliser, the respective modular form becomes an eigenvector of $\rho_I(\gamma)$ with $(c\tau_{\gamma}+d)^{-2k}$ as the eigenvalue. Making use of this characteristic, the alignment of modular forms at stabilisers are simple to obtain.

The stabilisers for the $A_5$ modular group are shown in TABLE \ref{table:stabilisers A5} and can be found in \cite{deMedeirosVarzielas:2020kji}. 
\begin{table}[ht]
\centering
\begin{tabular}{c|c} 
\toprule
$\gamma$ & $\tau_{\gamma}$\\ [0.5ex] 
\colrule
$T_{\tau}\,,\, T_{\tau}^2\,,\, T_{\tau}^3\,,\, T_{\tau}^4$ & $i\infty\,,\, \frac{8}{5}$ \\  
$S_{\tau}$ & $i\,,\, -\frac{70}{29}+\frac{i}{29}$ \\ 
$T_{\tau}S_{\tau}\,,\, T_{\tau}S_{\tau}T_{\tau}S_{\tau}$ & $\frac{1}{2}+\frac{i\sqrt{3}}{2}\,,\, -\frac{37}{26}+\frac{i}{26\sqrt{3}}$ \\
$S_{\tau}T_{\tau}\,,\, S_{\tau}T_{\tau}S_{\tau}T_{\tau}$ & $-\frac{1}{2}+\frac{i\sqrt{3}}{2}\,,\, \frac{91}{38}+\frac{i\sqrt{3}}{38}$ \\ [1ex]
\botrule
\end{tabular}
\caption{Stabilisers for some of the $A_5$ elements \cite{deMedeirosVarzielas:2020kji}.}
\label{table:stabilisers A5}
\end{table}

When considering $S_{\tau}$, $T_{\tau}$, $S_{\tau}T_{\tau}$ and $T_{\tau}S_{\tau}$, the respective $(c\tau_{\gamma}+d)^{-2k}$ are
\begin{equation}
(c \tau_{\gamma} + d)^{-2k} = 
\left\{
\begin{array}{ll}
      (-1)^{k} & \tau_{S_{\tau 1}} = i \\
      1 & \tau_{T_{\tau 1}} = i\infty \\
\end{array}
\right. .
\label{stabilisers eigenvalues A5}
\end{equation}

The alignments of the modular forms of weight $2k=2$ and $4$ for the $\tau$ that  stabilise the generators $S$ and $T$ are shown in 
TABLE \ref{table:directions A5}.
We present also  the associated factors in terms of $Y$, defined as the first component $Y_1$ of the modular form $Y_\bfqui^{(2)}$. We used the definitions for the modular forms of weight 2 present in Appendix \ref{secapp:modular forms A5}. We include also the value of the singlet modular form of weight 4.

\begin{table}[ht!]
\centering
{\renewcommand{\arraystretch}{1.4}
\begin{tabular}{c|c|c|c}
\toprule
\multicolumn{2}{c|}{$\tau_\gamma$} & $\tau_{S_{\tau 1}}=i$ & $\tau_{T_{\tau 1}}=i\infty$\\
\colrule
& $\bfqui$ & $Y \begin{pmatrix} 1 \\ 
\frac{-1-\sqrt{7-4\phi}}{\sqrt{6}} \\ 
\frac{-1-\sqrt{18-11\phi}}{\sqrt{6}} \\ 
\frac{-1+\sqrt{18-11\phi}}{\sqrt{6}} \\ \frac{-1+\sqrt{7-4\phi}}{\sqrt{6}} \end{pmatrix}$ & $Y \begin{pmatrix}1\\0\\0\\0\\0\end{pmatrix}$ \\
\cline{2-4}
weight 2 & $\bft$ & $Y \begin{pmatrix} \sqrt{\frac{58-31\phi}{15}} \\
\frac{-9+8\phi+\sqrt{27-4\phi}}{\sqrt{30}} \\ 
\frac{9-8\phi+\sqrt{27-4\phi}}{\sqrt{30}} \end{pmatrix}$ & $\sqrt{\frac{3}{5}}Y \begin{pmatrix}1\\0\\0\end{pmatrix}$ \\
\cline{2-4}
& $\bftp$ & $Y \begin{pmatrix} -\sqrt{\frac{3+4\phi}{15}} \\
\frac{7-4\phi+\sqrt{2+\phi}}{\sqrt{30}} \\ 
\frac{-7+4\phi+\sqrt{2+\phi}}{\sqrt{30}} \end{pmatrix}$ & $-\sqrt{\frac{3}{5}}Y \begin{pmatrix}1\\0\\0\end{pmatrix}$ \\
\colrule
& $\bfs$ & $\frac{15\sqrt{5}-25}{6} Y^2$ & $Y^2$ \\
\cline{2-4}
& $\bft$ & $ -\sqrt{\frac{100-40 \sqrt{5}}{3}} Y^2 \begin{pmatrix} 
1 \\
-\frac{\sqrt{3 - \sqrt{5}}}{2} \\
-\frac{\sqrt{3 - \sqrt{5}}}{2}
\end{pmatrix}$ & $-2\sqrt{\frac{3}{5}}Y^2 \begin{pmatrix}1\\0\\0\end{pmatrix}$ \\
\cline{2-4}
& $\bftp$ & $\sqrt{\frac{125-55\sqrt{5}}{2}} Y^2
\begin{pmatrix}
1 \\
\frac{\sqrt{3 + \sqrt{5}}}{2} \\
\frac{\sqrt{3 + \sqrt{5}}}{2}
\end{pmatrix}$ & $\frac{3}{\sqrt{5}}Y^2 \begin{pmatrix}1\\0\\0\end{pmatrix}$\\
\cline{2-4}
\multirow{2}{*}{weight 4} & $\bfqua$ & $\frac{Y^2}{12} \begin{pmatrix}
25-15 \sqrt{5}-5\sqrt{10-2 \sqrt{5}} \\
25-15 \sqrt{5}+5\sqrt{130-58 \sqrt{5}} \\
25-15 \sqrt{5}-5\sqrt{130-58 \sqrt{5}} \\
25-15 \sqrt{5}+5\sqrt{10-2 \sqrt{5}}
\end{pmatrix}$ & $0$\\
\cline{2-4}
& $\bfqui_1$ & $Y^2 \begin{pmatrix}
\frac{1}{6} \sqrt{15 \sqrt{5}+35} \\
\frac{-11 \sqrt{5}+2 \sqrt{250-110 \sqrt{5}}+35}{4 \sqrt{3}} \\
-\frac{-7 \sqrt{5}+2 \sqrt{5 \left(5-2 \sqrt{5}\right)}+15}{2 \sqrt{3}} \\
\sqrt{\frac{5}{3} \left(5-2 \sqrt{5}\right)}+\sqrt{\frac{5}{6} \left(47-21 \sqrt{5}\right)} \\
\frac{-11 \sqrt{5}-2 \sqrt{250-110 \sqrt{5}}+35}{4 \sqrt{3}}
\end{pmatrix}$ & $\sqrt{2}Y^2 \begin{pmatrix}1\\0\\0\\0\\0\end{pmatrix}$\\
\cline{2-4}
& $\bfqui_2$ & $Y^2 \begin{pmatrix}
-\frac{7-\sqrt{45}}{\sqrt{3}} \\
-\frac{1}{3} \sqrt{-173 \sqrt{5}+8 \sqrt{10-2 \sqrt{5}}+407} \\
\frac{2}{3} \sqrt{-61 \sqrt{5}+4 \sqrt{1930-862 \sqrt{5}}+143} \\
-\frac{2}{3} \sqrt{-61 \sqrt{5}-4 \sqrt{1930-862 \sqrt{5}}+143} \\
-\frac{1}{3} \sqrt{-173 \sqrt{5}-8 \sqrt{10-2 \sqrt{5}}+407}
\end{pmatrix}$ & 0\\
\colrule
\multicolumn{2}{c|}{$Y$} & $2.594\ldots i$ & $\sqrt{\frac{2}{3}} \pi i$ \\
\botrule
\end{tabular}
}
\caption{Directions for the modular forms of weight 2 and 4 of level 5 for the $A_5$ generators.}
\label{table:directions A5}
\end{table}

\section{Golden ratio mixing and related mixings}
\label{sec:GR mixing}

The golden ratio (GR) mixing is a mixing associated in previous works with models based in the $A_5$ symmetry,
and this is not different for models using multiple modular $A_5$. 
The mixing matrix that we will use is 
\begin{equation}
U_{GR}=
\begin{pmatrix}
\frac{\phi}{\sqrt{2+\phi}} & \frac{1}{\sqrt{2+\phi}} & 0 \\
-\frac{1}{\sqrt{4+2\phi}} & \frac{\phi}{\sqrt{4+2\phi}} & 1/\sqrt{2} \\
-\frac{1}{\sqrt{4+2\phi}} & \frac{\phi}{\sqrt{4+2\phi}} & -1/\sqrt{2}
\end{pmatrix},
\label{eq:GoldenRatio}
\end{equation}
where $\phi = \frac{1 + \sqrt{5}}{2}$.
This mixing has the same problem as the TBM mixing: it is incompatible with the experimental results for $\theta_{13}$, and thus we want to work with models that preserve only the first or the second columns of the GR mixing matrix, that can be written as the GR matrix times a rotation between the other two columns.

For a model where the second column is preserved, the matrix that diagonalizes $M_{\nu}$ is $U = U_{GR} U_r$, where $U_r$ is a rotation between the first and third columns. 
Using the parametrisation
\begin{equation}
U_r = \begin{pmatrix}
\cos\theta e^{i \alpha_1} & 0 & \sin\theta e^{-i \alpha_2} \\
0 & e^{i \alpha_3} & 0 \\
-\sin\theta e^{i \alpha_2} & 0 & \cos\theta e^{- i \alpha_1}
\end{pmatrix},
\label{eq:rotation13 for GR2}
\end{equation}
we are then able to diagonalize $M_{\nu}$. Here, $\theta$ is the angle that governs the rotation and the three $\alpha_i$ are introduced such that $m_i$ are purely real values.

The angles and phases from the standard parametrisation of the PMNS matrix in \cite{Zyla:2020zbs} can be expressed in terms of the model parameters $\theta$, $\alpha_1$ and $\alpha_2$ using the expressions between the parameters and the PMNS matrix elements:
\begin{align}
\label{eq:sintheta13 relation A5 GR2} \sin^2\theta_{13} &= |U_{e3}|^2  = \frac{5+\sqrt{5}}{10}\sin^2\theta \\
\label{eq:sintheta12 relation A5 GR2} \sin^2\theta_{12} &= \frac{|U_{e2}|^2}{1-|U_{e3}|^2} = \frac{3-\sqrt{5}}{4-\sqrt{5}+\cos 2\theta} \\
\label{eq:sintheta23 relation A5 GR2} \sin^2\theta_{23} &= \frac{|U_{\mu3}|^2}{1-|U_{e3}|^2} = \frac{4 - \sqrt{5} + \cos2\theta - 2\sqrt{5-2\sqrt{5}} \sin2\theta \cos (\alpha_1-\alpha_2)}{8 - 2\sqrt{5} + 2\cos 2\theta} \\
\delta &= - \arg \left( \frac{ U_{e3} U_{\tau 1} U_{e1}^* U_{\tau 3}^*}{\cos\theta_{12} \sin\theta_{13} \cos^2\theta_{13} \cos\theta_{23}} 
+ \cos\theta_{12} \sin\theta_{13} \cos\theta_{23} \right) \nonumber \\ 
\label{eq:delta relation A5 GR2} & = \arg \left(\sin2\theta \left( \frac{5 + \sqrt{5}}{2} e^{-i (\alpha_1-\alpha_2)} \cos^2\theta - e^{i (\alpha_1-\alpha_2)} \sin^2\theta \right) \right).
\end{align}

Using the $3\sigma$ C.L. range of $\sin^2\theta_{13}$ for NO(IO), $0.02034(0.02053)\rightarrow0.02430(0.02434)$ \cite{Esteban:2020cvm}, we obtain the allowed range for $\sin\theta$:
\begin{equation}
0.1677(0.1684)\lesssim|\sin\theta|\lesssim0.1833(0.1834),
\end{equation} 
which implies also ranges for the other mixing angles (using that $-1\leq\cos(\alpha_1-\alpha_2)\leq1$): 
\begin{align}
& 0.2821 (0.2822) \lesssim \sin^2\theta_{12} \lesssim 0.2833 (0.2833) \\
& 0.4029 (0.4028) \lesssim \sin^2\theta_{23} \lesssim 0.5971 (0.5972).
\end{align}
The $1\sigma$ NuFit region is within the interval found for $\sin^2\theta_{23}$, which overlaps with the $3\sigma$ region for this parameter, with our result extending below 0.405(0.410) for NO(IO) and not reaching its upper limit. 
The range of allowed values for $\sin^2\theta_{12}$ is near the lowest limit of the $1\sigma$ region although outside.

For a model where the first column is preserved instead, the rotation matrix $U_r$ between the second and third columns can be parametrised as:
\begin{equation}
U_r = \begin{pmatrix}
e^{i \alpha_3} & 0 & 0 \\
0 & \cos\theta e^{i \alpha_1} & \sin\theta e^{-i \alpha_2} \\
0 & -\sin\theta e^{i \alpha_2} & \cos\theta e^{- i \alpha_1}
\end{pmatrix},
\label{eq:rotation23 for GR1}
\end{equation}
Again, $\theta$ is the angle that governs the rotation and the three $\alpha_i$ are introduced such that the three neutrino masses $m_i$ have purely real values.

For this model, the expressions for the angles and phases from the standard parametrisation of the PMNS matrix in \cite{Zyla:2020zbs} in terms of the model parameters $\theta$, $\alpha_1$ and $\alpha_2$ are
\begin{align}
\label{eq:sintheta13 relation A5 GR1} \sin^2\theta_{13} &= |U_{e3}|^2  = \frac{5-\sqrt{5}}{10}\sin^2\theta \\
\label{eq:sintheta12 relation A5 GR1} \sin^2\theta_{12} &= \frac{|U_{e2}|^2}{1-|U_{e3}|^2} = \frac{2\cos^2\theta}{4+\sqrt{5}+\cos 2\theta} \\
\label{eq:sintheta23 relation A5 GR1} \sin^2\theta_{23} &= \frac{|U_{\mu3}|^2}{1-|U_{e3}|^2} = \frac{4 + \sqrt{5} + \cos2\theta + 2\sqrt{5+2\sqrt{5}} \sin2\theta \cos (\alpha_1-\alpha_2)}{8 + 2\sqrt{5} + 2\cos 2\theta} \\
\delta &= - \arg \left( \frac{ U_{e3} U_{\tau 1} U_{e1}^* U_{\tau 3}^*}{\cos\theta_{12} \sin\theta_{13} \cos^2\theta_{13} \cos\theta_{23}} 
+ \cos\theta_{12} \sin\theta_{13} \cos\theta_{23} \right) \nonumber \\ 
\label{eq:delta relation A5 GR1} & = \arg \left(\sin2\theta \left( \frac{5 - \sqrt{5}}{2} e^{-i (\alpha_1-\alpha_2)} \cos^2\theta - e^{i (\alpha_1-\alpha_2)} \sin^2\theta \right) \right).
\end{align}

Using the $3\sigma$ C.L. range of $\sin^2\theta_{13}$ for NO(IO), $0.02034(0.02053)\rightarrow0.02430(0.02434)$ \cite{Esteban:2020cvm}, we obtain the allowed range for $\sin\theta$:
\begin{equation}
0.2713(0.2725)\lesssim|\sin\theta|\lesssim0.2965(0.2968),
\end{equation} 
which implies also ranges for the other mixing angles  (using that $-1\leq\cos(\alpha_1-\alpha_2)\leq1$): 
\begin{align}
& 0.2584 (0.2583) \lesssim \sin^2\theta_{12} \lesssim 0.2614 (0.2612) \\
& 0.2531 (0.2529) \lesssim \sin^2\theta_{23} \lesssim 0.7469 (0.7471).
\end{align}
We conclude that the range of allowed values for $\sin^2\theta_{12}$ is outside the $3\sigma$ region and thus the class of models that preserve the first column of the golden ratio mixing matrix, which we call GR$_1$ mixing, are disfavoured by experiment. 

Consequently, in the following we are only interested in models that preserve the second column of the golden ratio mixing, which we call GR$_2$, although, as pointed out previously, even for these models $\sin^2\theta_{12}$ is outside the experimental $1\sigma$ interval.

\section{Model with two modular \texorpdfstring{$A_5$}{A5} symmetries using the Weinberg operator} 
\label{sec:models weinberg A5}

Now that the $A_5$ modular symmetry and the mixing derived from the GR mixing were introduced, the model that uses this symmetry in order to get what we called the GR$_2$ mixing can now be described, assuming that neutrinos get their mass through the Weinberg operator.
At high energies, these models are based in two modular symmetries, $A_5^l$ and $A_5^{\nu}$, with modulus fields denoted by $\tau_l$ and $\tau_{\nu}$, respectively. 
After the modulus fields acquire different VEV's, different mass textures are realised in the charged lepton and neutrino sectors, in such a way that the GR$_2$ mixing is recovered for the PMNS. 

We consider then that neutrinos get their mass through an effective term of the type $\frac{1}{\Lambda} Y L^2 H_u^2$.
The transformation properties of fields and Yukawa couplings can be found in 
TABLE \ref{tab:transformationp weinbergL3b A5}. 

\begin{table}[htb]
\centering
{\renewcommand{\arraystretch}{1.3}
\begin{tabular}[t]{c||c|c|c|c|c}
\toprule
Fields & $SU(2)$ & $A_5^l$ & $A_5^{\nu}$ & $2k_l$ & $2k_{\nu}$ \\ \colrule
$L$ & $\mathbf{2}$ & $\bftb$ & $\bfs$ & $0$ & $+2$ \\ 
$E^c$ & $\mathbf{2}$ & $\bftb$ & $\bfs$ & $+4$ & $-2$ \\
$H_{u,d}$ & $\mathbf{2}$ & $\bfs$ & $\bfs$ & $0$ & $0$ \\  
$\Phi$ & $\bfs$ & $\bfqui$ & $\bfqui$ & $0$ & $0$ \\ 
\botrule
\end{tabular}}~~
{\renewcommand{\arraystretch}{1.3}
\begin{tabular}[t]{c||c|c|c|c}
\toprule
Yukawas & $A_5^l$ & $A_5^{\nu}$ & $2k_l$ & $2k_{\nu}$ \\ \colrule
$Y^l_\bfs$ & $\bfs$ & $\bfs$ & $+4$ & $0$ \\
$Y^l_{\bftb}$ & $\bftb$ & $\bfs$ & $+4$ & $0$ \\ 
$Y^l_\bfqui$ & $\bfqui$ & $\bfs$ & $+4$ & $0$ \\ 
$Y^\nu_\bfs$ & $\bfs$ & $\bfs$ & $0$ & $+4$ \\
$Y^\nu_{\bfqui_1}$ & $\bfs$ & $\bfqui$ & $0$ & $+4$ \\
$Y^\nu_{\bfqui_2}$ & $\bfs$ & $\bfqui$ & $0$ & $+4$ \\ 
\botrule
\end{tabular}}
\caption{Transformation properties of fields and Yukawa couplings for model using the Weinberg operator and two modular $A_5$.}
\vspace{-2mm}
\label{tab:transformationp weinbergL3b A5}
\end{table}

All the Yukawa coefficients $Y^l$ and $Y^\nu$ are modular forms of weight 4.
The right-handed lepton fields $E^c$ are arranged as a triplet $\bft$ or $\bftp$ of $A_5^l$ and singlets $\bfs$ of $A_5^{\nu}$, with weights $2k_l=+4$ and $2k_{\nu}=-2$. 
Similarly, the lepton doublets $L$ transform as a $\bftb$ of $A_5^l$ and a $\bfs$ of $A_5^{\nu}$, with weights $2k_l=0$ and $2k_{\nu}=+2$.
These are the correct choices for the weights such that the modular forms and fields in each term sum up to zero since the weight for the fields is not $2k$, which are the values that were introduced in this section, but $-2k$ instead.
$H_d$ and $H_u$ are the usual Higgs and an additional Higgs doublet as required in supersymmetric models. 
A bi-quintuplet $\Phi$, which is a quintuplet under both $A_5^l$ and $A_5^\nu$, is introduced.

The multiplication of two triplets has the decomposition $\bftb \otimes \bftb = \bfs \oplus \bftb \oplus \bfqui$, where the $\bftb$ component is antisymmetric. This means that $L^2$ only decomposes as $\bfs\otimes\bfqui$, and so it must combine with a singlet or quintuplet. This implies that we have only to consider the contributions from $Y^{\nu}_\bfs$, $Y^{\nu}_{\bfqui_1}$ and $Y^{\nu}_{\bfqui_2}$, each associated with a different complex constant $g_i$. For $Y^\nu$, we only consider the contribution from $\bfqui_1$ since the other weight 4 $\bfqui_2$ will vanish at the chosen stabiliser for $\tau_\nu$ as is shown below.

With the fields assigned in this manner, the superpotential for this model, 
which can be separated into one part containing the mass terms for the charged leptons and the other the neutrino mass terms, 
has the following form: 
\begin{align}
w &= w_e+w_{\nu},\\ 
w_e &= \left( \alpha_1 Y^l_\bfs(\tau_l)(L E^c)_\bfs 
+ \alpha_2 Y^l_{\bftb}(\tau_l) (L E^c)_{\bftb} 
+ \alpha_3 Y^l_\bfqui(\tau_l) (L E^c)_\bfqui \right) H_d,\\
w_{\nu} &= \frac{1}{\Lambda} L^2 \left[ Y^\nu_\bfs(\tau_\nu) + \frac{1}{\Lambda}
\Phi \left( Y^\nu_{\bfqui_1}(\tau_\nu) + Y^\nu_{\bfqui_2}(\tau_\nu) \right)\right] H_u^2.
\end{align}

\subsection{\texorpdfstring{$A_5^l \times A_5^{\nu} \rightarrow A_5^D$ breaking}{A5l x A5nu breaking to A5D}} 
\label{sec:symmetry breaking weinberg A5}

Considering the multiplication rules for two quintuplets to get a trivial singlet, the term $\frac{1}{\Lambda^2} L^2 \Phi Y^{\nu} H_u^2 $ can be explicitly expanded as:
\begin{equation}
\frac{1}{\Lambda^2}(L^2)_\bfqui^T P_{\pi}
\Phi P_{\pi} Y^{\nu}_\bfqui(\tau_{\nu}) H_u^2,
\label{eq:term with Phi weinberg A5}
\end{equation}
where $P_{\pi}$ is the matrix that describes the permutation
\begin{equation}
\pi = 
\begin{pmatrix}
1 & 2 & 3 & 4 & 5 \\
1 & 5 & 4 & 3 & 2 \\
\end{pmatrix},
\end{equation}
which is explicitly
\begin{equation}
P_\pi
=
\begin{pmatrix}
1 & 0 & 0 & 0 & 0 \\
0 & 0 & 0 & 0 & 1 \\
0 & 0 & 0 & 1 & 0 \\
0 & 0 & 1 & 0 & 0 \\
0 & 1 & 0 & 0 & 0 \\
\end{pmatrix}.
\end{equation}

If $\Phi$ acquires the VEV $\langle\Phi\rangle = v_{\Phi} P_{\pi}$ (see Appendix \ref{secapp:vac aligns A5 bi-qui} for more details), we get for the term in Eq.(\ref{eq:term with Phi weinberg A5})
\begin{equation}
\frac{v_\Phi}{\Lambda^2}(L^2)_\bfqui^T P_{\pi} Y^{\nu}_\bfqui(\tau_{\nu}) H_u^2,
\label{eq:term without Phi weinberg A5}
\end{equation}
which implies that $w_\nu$ gets the form (the $w_e$ terms remain exactly the same):
\begin{align}
w_{\nu} &= \frac{1}{\Lambda} \left[
(L^2)_\bfs Y^\nu_\bfs(\tau_\nu) + \frac{v_{\Phi}}{\Lambda} \left( (L^2)_\bfqui Y^\nu_{\bfqui_1}(\tau_\nu) + (L^2)_\bfqui Y^\nu_{\bfqui_2}(\tau_\nu) \right)_\bfs
\right] H_u^2.
\end{align}
This means that the symmetry $A_5^l\times A_5^{\nu}$ is broken but given that the same transformation $\gamma$ can be performed in $A_5^l$ and $A_5^{\nu}$ simultaneously and being the terms in the superpotential above all left invariant by such a transformation, there is still a single modular symmetry $A_5^D$, the diagonal subgroup, that is conserved.

The superpotential above implies a neutrino mass matrix. Expanding $Y^\nu_{\bfqui_1}$ and $Y^\nu_{\bfqui_2}$ in terms of the weight 2 modular functions gives the results already derived in \cite{Novichkov:2018nkm}.
If the triplets $L$, $E^c$ and $\nu^c$ are triplets $\bft$, which we will simply write as $\rho_L \sim \bft$, the neutrino mass matrix after the Higgs field acquires the VEV $\langle H_u \rangle = (0,v_u)$ gets the form:
\begin{align}
M_\nu^{\bft} &= 
g_\bfs (Y_1^2 + 2Y_3Y_4 + 2Y_2Y_5)
\begin{pmatrix}
1&0&0\\0&0&1\\0&1&0
\end{pmatrix} \nonumber\\
& + g_{\bfqui_1}
\begin{pmatrix}
Y_5 Y_7 - Y_2 Y_8 & -\frac{1}{2} Y_5 Y_6 - {\frac{1}{\sqrt2}} Y_4 Y_7 - \frac{\sqrt3}{2} Y_1 Y_8 & \frac{1}{2} Y_2 Y_6 + \frac{1}{\sqrt2} Y_3 Y_8 + \frac{\sqrt3}{2} Y_1 Y_7 \\
\ast & Y_5 Y_8 + \sqrt{2} Y_4 Y_6 & - \frac{1}{2} Y_5 Y_7 + \frac{1}{2} Y_2 Y_8 
\\
\ast & \ast & - Y_2 Y_7 -\sqrt{2}Y_3 Y_6
\end{pmatrix} \nonumber \\
& + g_{\bfqui_2}
\begin{pmatrix}
Y_1^2 + Y_3Y_4 - 2Y_2Y_5  & -\frac{3}{2\sqrt{2}} Y_3^2 + \sqrt{3} Y_1 Y_5 &
-\frac{3}{2\sqrt{2}} Y_4^2 + \sqrt{3} Y_1 Y_2 \\
\ast & 3Y_2 Y_3+\sqrt{\frac{3}{2}}Y_1 Y_4  &
- \frac{1}{2} Y_1^2 - \frac{1}{2}Y_3 Y_4 + Y_2 Y_5 \\
\ast & \ast & 3Y_4 Y_5 + \sqrt{\frac{3}{2}}Y_1 Y_3  
\end{pmatrix},
\end{align}
where asterisks were used to omit the off diagonal entries of symmetric matrices and $g_\bfs$, $g_{\bfqui_1}$ and $g_{\bfqui_2}$ are arbitrary complex constants associated with the respective modular form contribution. 
The factors $2 v_u^2/\Lambda$ and $2 v_u^2 v_\Phi/\Lambda^2$ are included inside these constants.

If the triplets $L$, $E^c$ and $\nu^c$ are triplets $\bftp$ instead, which can be equivalently expressed as $\rho_L \sim \bftp$, one obtains:
\begin{align}
M_\nu^{\bftp} &= 
g_\bfs
(Y_1^2 + 2Y_3Y_4 + 2Y_2Y_5)
\begin{pmatrix}
1&0&0\\0&0&1\\0&1&0
\end{pmatrix} \nonumber \\
& + g_{\bfqui_1}
\begin{pmatrix}
Y_5 Y_7-Y_2 Y_8 & - Y_4 Y_6 - \frac{1}{\sqrt{2}} Y_5 Y_8 & Y_3 Y_6 + \frac{1}{\sqrt{2}} Y_2 Y_7 \\
\ast  & - Y_3 Y_8 - \frac{1}{\sqrt{2}} Y_2 Y_6 - \sqrt{\frac{3}{2}} Y_1 Y_7 & - \frac{1}{2} Y_5 Y_7 + \frac{1}{2} Y_2 Y_8 \\
\ast & \ast & Y_4 Y_7 + \frac{1}{\sqrt{2}} Y_5 Y_6 + \sqrt{\frac{3}{2}} Y_1 Y_8
\end{pmatrix} \nonumber \\
& + g_{\bfqui_2} 
\begin{pmatrix}
Y_1^2 + Y_3Y_4-2Y_2Y_5  & -\frac{3}{\sqrt{2}}Y_2 Y_3 - \frac{\sqrt{3}}{2}Y_1 Y_4 & - \frac{3}{\sqrt{2}} Y_4 Y_5 - \frac{\sqrt{3}}{2}Y_1 Y_3 \\
\ast & \frac{3}{2} Y_4^2 - \sqrt{6} Y_1 Y_2 & - \frac{1}{2} Y_1^2 - \frac{1}{2} Y_3 Y_4 + Y_2 Y_5 \\
\ast & \ast & \frac{3}{2} Y_3^2 - \sqrt{6} Y_1 Y_5
\end{pmatrix},
\end{align}
where again $g_\bfs$, $g_{\bfqui_1}$ and $g_{\bfqui_2}$ are arbitrary complex constants associated with the respective modular form contribution that absorbed the factors $2 v_u^2/\Lambda$ and $2 v_u^2 v_\Phi/\Lambda^2$.

\subsection{\texorpdfstring{$A_5^D$ breaking}{A5D breaking}} 
\label{sec:A5D symmetry breaking weinberg A5}

The flavour structure after $A_5^D$ symmetry breaking will now be covered.
We assume that the charged lepton modular field $\tau_l$ acquires the VEV $\langle\tau_l\rangle = \tau_T = i\infty$.
This is a stabiliser of $T_\tau$ which means that a residual modular $Z_5^T$ symmetry is preserved in the charged lepton sector. 
The directions the modular forms take at this stabiliser are in 
TABLE \ref{table:directions A5}.
These directions lead to an almost diagonal charged lepton mass matrix when the Higgs field $H_d$ acquires a VEV $\langle H_d \rangle = (0,v_d)$:
\begin{equation}
m_e=v_d\alpha_1\begin{pmatrix}
1+2\frac{\alpha_3}{\alpha_1} & 0 & 0 \\
0 & 0 & 1-\frac{\alpha_2}{\alpha_1}-\frac{\alpha_3}{\alpha_1} \\
0 & 1+\frac{\alpha_2}{\alpha_1}-\frac{\alpha_3}{\alpha_1} & 0
\end{pmatrix}.
\label{mass matrix charged leptons weinberg A5}
\end{equation}
The masses for the charged leptons can be reproduced by adjusting the parameters $\alpha_i$. 
These constants were redefined to include the constant associated with $Y^l(\tau_l)$.
This matrix can be diagonalized by multiplying on the left and right by $P_L$ and $P_R$ ($P_L^T m_e P_R=m_{e_d}$) by taking $P_L$ as the identity matrix and $P_R=P_{23}$. 
Consequently, the PMNS matrix is simply the matrix that diagonalizes the mass matrix for the neutrinos.
These considerations are valid whether we choose the triplets in the model to be $\bft$ or $\bftp$.

For the other modular field $\tau_{\nu}$, we want to find a VEV that leads to a mixing that preserves the second column of the GR mixing matrix. 
This occurs for $\langle \tau_{\nu} \rangle = \tau_S = i$ and for $Y^\nu$ with weight 4 (see 
TABLE \ref{table:directions A5} 
for the directions the modular forms get at this stabiliser). 
In this case, a residual modular $Z_2^S$ symmetry is preserved in the neutrino sector.

If $\rho_L \sim \bft$, this implies the following structure for the neutrino mass matrix:

\vspace{-4mm}
{\small
\begin{align}
M_\nu^\bft &=
g_\bfs 
\left(
\begin{array}{ccc}
 1 & 0 & 0 \\
 0 & 0 & 1 \\
 0 & 1 & 0 \\
\end{array}
\right) \nonumber \\
& + g_{\bfqui_1}
\left(
\begin{array}{ccc}
 1 & - \sqrt{ \frac{241}{8} + 13 \sqrt{5} - \sqrt{1525 + 682 \sqrt{5}} } & - \sqrt{ \frac{241}{8} + 13 \sqrt{5} + \sqrt{1525 + 682 \sqrt{5}} } \\
 \ast & - 3 - 2 \sqrt{5} + \sqrt{50 + 22 \sqrt{5}} & -\frac{1}{2} \\
 \ast & \ast & - 3 - 2 \sqrt{5} - \sqrt{50+22 \sqrt{5}} \\
\end{array}
\right) \nonumber \\
& + g_{\bfqui_2} 
\left(
\begin{array}{ccc}
 1 & - \frac{3}{2} \sqrt{ \frac{949}{2} - 212 \sqrt{5} - 2 \sqrt{ 103 445 - 46 262 \sqrt{5}} } & - \frac{3}{2} \sqrt{ \frac{949}{2} - 212 \sqrt{5} + 2 \sqrt{ 103 445 - 46 262 \sqrt{5}} }  \\
 \ast & \frac{3}{2} \left( 18 - 8 \sqrt{5} + \sqrt{ 130 - 58 \sqrt{5} } \right) & -\frac{1}{2} \\
 \ast & \ast & \frac{3}{2} \left( 18 - 8 \sqrt{5} - \sqrt{ 130 - 58 \sqrt{5} } \right) \\
\end{array}
\right)
\nonumber \\
& \approx \left(
\begin{array}{ccc}
 g_\bfs + g_{\bfqui_1} + g_{\bfqui_2} & -1.99176 g_{\bfqui_1} - 0.578608 g_{\bfqui_2} & -10.6968 g_{\bfqui_1} - 1.30628 g_{\bfqui_2}\\
 \ast & 2.48746 g_{\bfqui_1} + 0.999728 g_{\bfqui_2} & g_\bfs - \frac{1}{2} g_{\bfqui_1} -\frac{1}{2} g_{\bfqui_2} \\
 \ast & \ast & -17.4317 g_{\bfqui_1} - 0.665359 g_{\bfqui_2} \\
\end{array}
\right)
\end{align}
}%
where $g_\bfs$, $g_{\bfqui_1}$ and $g_{\bfqui_2}$ were redefined to include factors coming from the modular forms $Y^\nu_\bfs$, $Y^\nu_{\bfqui_1}$ and  $Y^\nu_{\bfqui_2}$.

We want now to diagonalize $M_{\nu}$, such that $U^T M_{\nu} U = M_{\nu_d} = \text{diag}(m_1,m_2,m_3)$, where $m_i$ are the neutrino masses and $U$ is an unitary matrix.
When we apply the golden ratio mixing matrix Eq.(\ref{eq:GoldenRatio}) to the neutrino mass matrix for triplets $\bft$ one obtains: 
\begin{equation}
U_{GR}^T M_\nu^\bft U_{GR} = 
\left(
\begin{array}{ccc}
 \frac{1}{10} \left(\left(7 \sqrt{5} + 5\right) a +\left(7 \sqrt{5}-5\right) b +16 \sqrt{5} c \right) & 0 & 0  \\
 0 & a & c \\
 0 & c & b  \\
\end{array}
\right)
\label{eq:UMnuU1 weinbergL3 A5}
\end{equation}
where $a = g_\bfs - \frac{13 \sqrt{5}+25}{4} g_{\bfqui_1} + \frac{27 \sqrt{5}-65}{4} g_{\bfqui_2}$, $b = -2 g_\bfs - \frac{4 \sqrt{5} + 5}{2} g_{\bfqui_1} + \frac{55 - 24 \sqrt{5}}{2}  g_{\bfqui_2}$ and $c = \left( 3 \sqrt{5} + 5 \right) g_{\bfqui_1} + \frac{3 \left( 7 \sqrt{5} - 15 \right)}{2}  g_{\bfqui_2}$.

This implies that the PMNS is simply the Golden Ratio matrix times a rotation among the second and third columns, conserving only its first column. We have already discussed the compatibility of the GR$_1$ mixing and experimental values in Section \ref{sec:GR mixing}, where it has already been seen that this mixing is incompatible with the $3\sigma$ confidence interval for $\theta_{12}$. For this reason, we will not further develop the case $\rho_L \sim \bft$.

We now turn our attention to $M_\nu^\bftp$.
For $\rho_L \sim \bftp$, we have the following structure for the neutrino mass matrix:

\vspace{-4mm}
{\small
\begin{align}
M_\nu^\bftp &= 
g_\bfs 
\left(
\begin{array}{ccc}
 1 & 0 & 0 \\
 0 & 0 & 1 \\
 0 & 1 & 0 \\
\end{array}
\right) \nonumber \\
& + g_{\bfqui_1} 
\left(
\begin{array}{ccc}
 1 & -\sqrt{\frac{79}{2} + 17 \sqrt{5}-\sqrt{2770 + 1238 \sqrt{5}}} &
 \sqrt{\frac{79}{2} + 17 \sqrt{5} + \sqrt{2770 + 1238 \sqrt{5}}} \\
 \ast & \frac{9}{2} + 2 \sqrt{5} + 2 \sqrt{5 + 2 \sqrt{5}} & - \frac{1}{2} \\
 \ast & \ast & \frac{9}{2} + 2 \sqrt{5} - 2 \sqrt{5 + 2 \sqrt{5}} \\
\end{array}
\right) \nonumber \\
& + g_{\bfqui_2} 
\left(
\begin{array}{ccc}
 1 & - \frac{3}{2} \sqrt{387 - 173 \sqrt{5} + 2 \sqrt{41810 - 18698 \sqrt{5}} } &
 \frac{3}{2} \sqrt{387 - 173 \sqrt{5} - 2 \sqrt{41810 - 18698 \sqrt{5}}} \\
 \ast & -\frac{51}{2} + 12 \sqrt{5} + 3 \sqrt{85 - 38 \sqrt{5}} & - \frac{1}{2}  \\
 \ast & \ast & -\frac{51}{2} + 12 \sqrt{5} - 3 \sqrt{85-38 \sqrt{5}} \\
\end{array}
\right) \nonumber \\
&\approx \left(
\begin{array}{ccc}
 g_\bfs + g_{\bfqui_1} + g_{\bfqui_2} & -1.75890 g_{\bfqui_1} - 0.706914 g_{\bfqui_2} & 12.3261 g_{\bfqui_1} + 0.47048 g_{\bfqui_2} \\
 \ast & 15.1275 g_{\bfqui_1} + 1.84736 g_{\bfqui_2} & g_\bfs + 0.5 g_{\bfqui_1} + 0.5 g_{\bfqui_2} \\
 \ast & \ast & 2.81677 g_{\bfqui_1} + 0.818275 g_{\bfqui_2} \\
\end{array}
\right)
\end{align}
}%
where once again $g_\bfs$, $g_{\bfqui_1}$ and $g_{\bfqui_2}$ were redefined to include the factors coming from the modular forms $Y^\nu_\bfs$, $Y^\nu_{\bfqui_1}$ and $Y^\nu_{\bfqui_2}$.

When we apply the golden ratio mixing matrix Eq.(\ref{eq:GoldenRatio}) to the neutrino mass matrix for triplets $\bftp$ we obtain:
\begin{equation}
U_{GR}^T M_\nu^\bftp U_{GR} = 
\begin{pmatrix}
 a & 0 & c  \\
 0 & \frac{1}{10} \left( \left( 5 - \sqrt{5} \right) a - \left( 5 + \sqrt{5} \right) b - 8 \sqrt{5} c \right) & 0 \\
 c & 0 & b  \\
\end{pmatrix},
\label{eq:UMnuU1 weinbergL3p A5}
\end{equation}
where $a = g_\bfs - \frac{5 + \sqrt{5}}{2} g_{\bfqui_1} + \frac{39 \sqrt{5}-85}{2} g_{\bfqui_2}$, $b = - g_\bfs + \left( 5 + 2 \sqrt{5} \right) g_{\bfqui_1} + \left( 12 \sqrt{5} - 25 \right) g_{\bfqui_2}$ and $c = - \left(5 + 3 \sqrt{5} \right) g_{\bfqui_1}$ $- \frac{3}{2} \left( 7 \sqrt{5}-15 \right) g_{\bfqui_2}$.
This matrix has only an element on the second row and second column and four elements on the corners that form a $2\times2$ symmetric matrix and so it can be fully diagonalized by introducing a matrix $U_r$ that describes a rotation among the first and third columns. 
The matrix that diagonalizes $M_{\nu}$ is then $U = U_{GR} U_r$, where $U_r$ is given by Eq.(\ref{eq:rotation13 for GR2}). 
We are then able to diagonalize $M_{\nu}$ and the lepton mixing obeys a GR$_2$ mixing.

It is also possible to start from the diagonal matrix $M_{\nu_d}$ and get $U_{GR}^T M_\nu U_{GR}$. We have that:
\begin{equation}
\resizebox{.9\hsize}{!}{
$U_r^* M_{\nu_d} U_r^{\dagger} =
\begin{pmatrix}
m_1 e^{-2 i \alpha_1} \cos^2\theta + m_3 e^{2 i \alpha_2} \sin^2\theta & 0 & \frac{1}{2} ( - m_1 e^{-i (\alpha_1 + \alpha_2)} + m_3 e^{i(\alpha_1 + \alpha_2)}) \sin2\theta \\
0 & m_2 e^{-2i\alpha_3} & 0\\
\ast & 0 & m_1 e^{-2 i \alpha_2} \sin^2\theta + m_3e^{2 i \alpha_1} \cos^2\theta
\end{pmatrix}$}
\label{eq:UMnudU2 weinbergL3p A5},
\end{equation}
and comparing with (\ref{eq:UMnuU1 weinbergL3p A5}) we obtain that
$\alpha_3 =  - \frac{1}{2} \arg\left( \left( 5 - \sqrt{5} \right) a - \left( 5 + \sqrt{5} \right) b - 8 \sqrt{5} c \right)$ and, more importantly, we get a mass sum rule for $m_i$:
\begin{align}
m_2 & = \Big| \frac{1}{10} \left( \left( 5 - \sqrt{5} \right) a - \left( 5 + \sqrt{5} \right) b - 8 \sqrt{5} c \right) \Big| \nonumber \\
\label{eq:sum rule weinbergL3p A5} & = \frac{1}{10} \Big| m_1 \left( \left(5 - \sqrt{5}\right) e^{-2 i \alpha_1} \cos^2\theta - \left(5 + \sqrt{5}\right) e^{-2 i \alpha_2} \sin^2\theta + 4 \sqrt{5} e^{-i (\alpha_1+\alpha_2)} \sin 2 \theta \right) - \\
& \hspace{1.8cm} - m_3 \left( \left(5 + \sqrt{5}\right) e^{2 i \alpha_1} \cos^2\theta - \left(5 - \sqrt{5}\right) e^{2 i \alpha_2} \sin^2\theta + 4 \sqrt{5} e^{i (\alpha_1+\alpha_2)} \sin 2 \theta \right) \Big|.
\nonumber
\end{align}

The sum rule (\ref{eq:sum rule weinbergL3p A5}) and (\ref{eq:sintheta13 relation A5 GR2}-\ref{eq:delta relation A5 GR2}) give us relations between the six observables (the three mixing angles, the atmospheric and solar neutrino squared mass differences and the Dirac neutrino CP violation phase) and the five parameters of the GR$_2$ mixing ($\theta$, $\alpha_1$, $\alpha_2$, $m_1$ and $m_2$), and hence we can do a numerical minimisation using the $\chi^2$ function:
\begin{equation}
\chi^2 = \sum_i \left(\frac{P_i(\{x\})-BF_i}{\sigma_i}\right)^2.
\label{eq:chisq A5}
\end{equation}
$P_i$ are the model values, $BF$ are the best fit values from NuFit \cite{Esteban:2020cvm}, $\sigma_i$ is obtained by averaging the upper and lower $\sigma$ provided by NuFit.

The fit parameters obtained for normal ordering (NO) and inverted ordering (IO) of neutrino masses can be found in 
TABLE \ref{tab:fitresults weinbergL3p A5}. 
The best fit values lie inside the 1$\sigma$ range for all the observables except $\theta_{12}$, for both orderings near the lower limit of the 1$\sigma$ range, and $\theta_{23}$ for IO. Nonetheless, all the observables are within their $3\sigma$ intervals. The best-fit occurs for NO with a $\chi^2=3.22$. 
\begin{table}[ht]
\centering
{\renewcommand{\arraystretch}{1.3}
\begin{tabular}{| c | c | c c c c c c c |}
\cline{1-9}
\multirow{4}{*}{NO} & \multirow{2}{*}{Para.} & & $\chi^2$ & $\theta$ & $\alpha_1$ & $\alpha_2$ & $m_1$ & $m_3$ \\
\cline{3-9}
& & & 3.22 & -10.09\degree & -22.77\degree & 8.72\degree & 0.0319 eV & 0.0595 eV \\
\cline{2-9}
& \multirow{2}{*}{Obs.} & $\theta_{12}$ & $\theta_{23}$ & $\theta_{13}$ & $\delta$ & $\Delta m_{21}^2$ & $\Delta m_{31}^2$ & $m_{\beta\beta}$\\
\cline{3-9}
& & 32.12\degree & 49.5\degree & 8.57\degree & 212\degree & 7.42$\times 10^{-5}$eV$^2$ & 2.515$\times 10^{-3}$eV$^2$ & 0.0276 eV\\
\cline{1-9}
\multirow{4}{*}{IO} &\multirow{2}{*}{Para.} & & $\chi^2$ & $\theta$ & $\alpha_1$ & $\alpha_2$ & $m_1$ & $m_3$ \\
\cline{3-9}
& & & 11.6 & -10.15\degree & -114.29\degree & -41.16\degree & 0.1288 eV & 0.1191 eV \\
\cline{2-9}
& \multirow{2}{*}{Obs.} & $\theta_{12}$ & $\theta_{23}$ & $\theta_{13}$ & $\delta$ & $\Delta m_{21}^2$ & $\Delta m_{32}^2$ & $m_{\beta\beta}$\\
\cline{3-9}
& & 32.12\degree & 46.6\degree & 8.62\degree & 253\degree & 7.42$\times 10^{-5}$eV$^2$ & -2.498$\times 10^{-3}$eV$^2$ & 0.1160 eV \\
\cline{1-9}
\end{tabular}}
\caption{Parameters (Para.) and observables (Obs.) for the best fit point for normal and inverted orderings for model using the Weinberg operator and two modular $A_5$.}
\label{tab:fitresults weinbergL3p A5}
\end{table}
\begin{figure}[ht]
     \centering
     \begin{subfigure}[t]{0.49\textwidth}
         \centering
         \includegraphics[width=\textwidth]{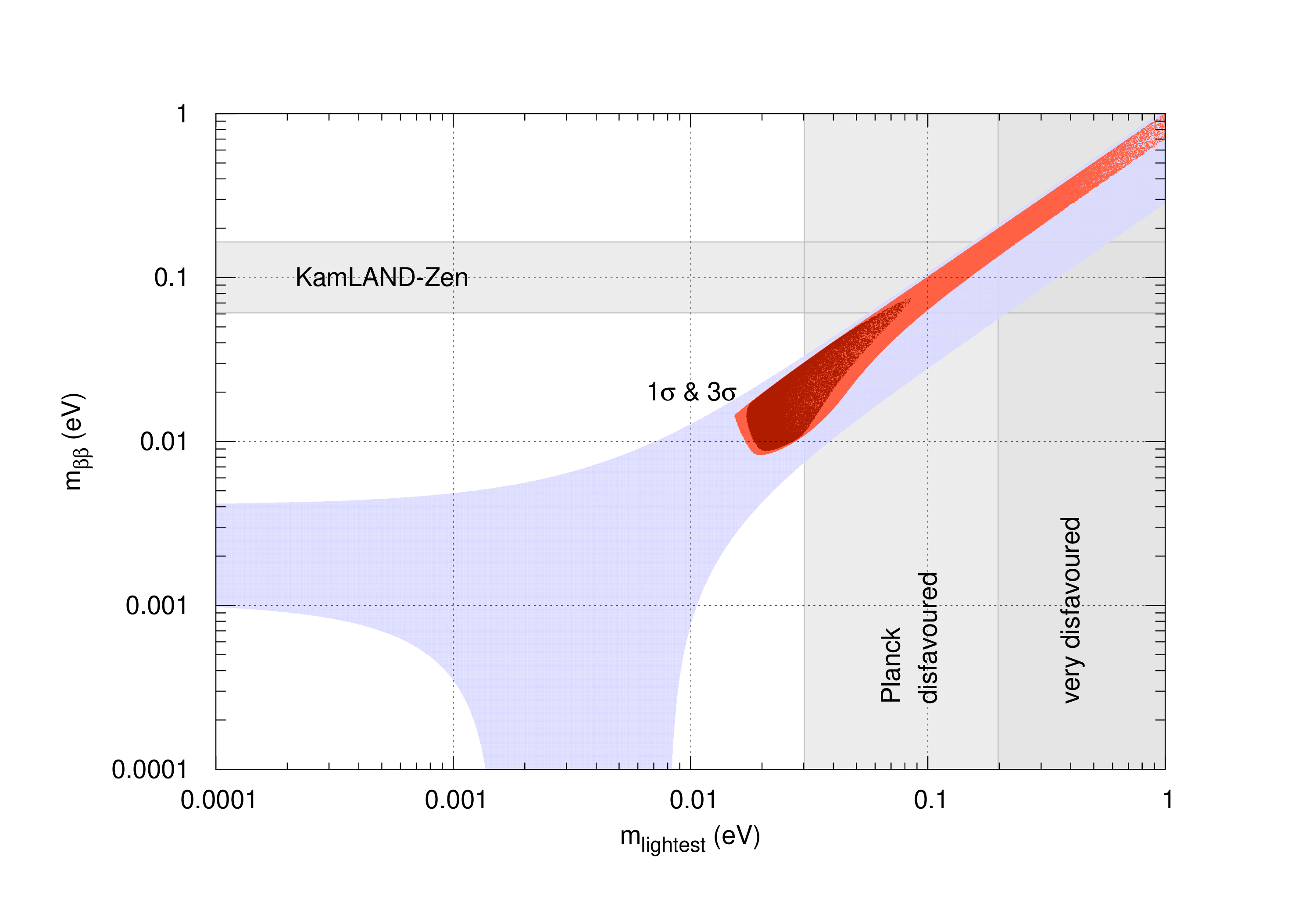}
		 \caption{Normal Ordering}
		 \label{fig:mee weinbergL3p NO A5}
     \end{subfigure}%
     ~
     \begin{subfigure}[t]{0.49\textwidth}
         \centering
         \includegraphics[width=\textwidth]{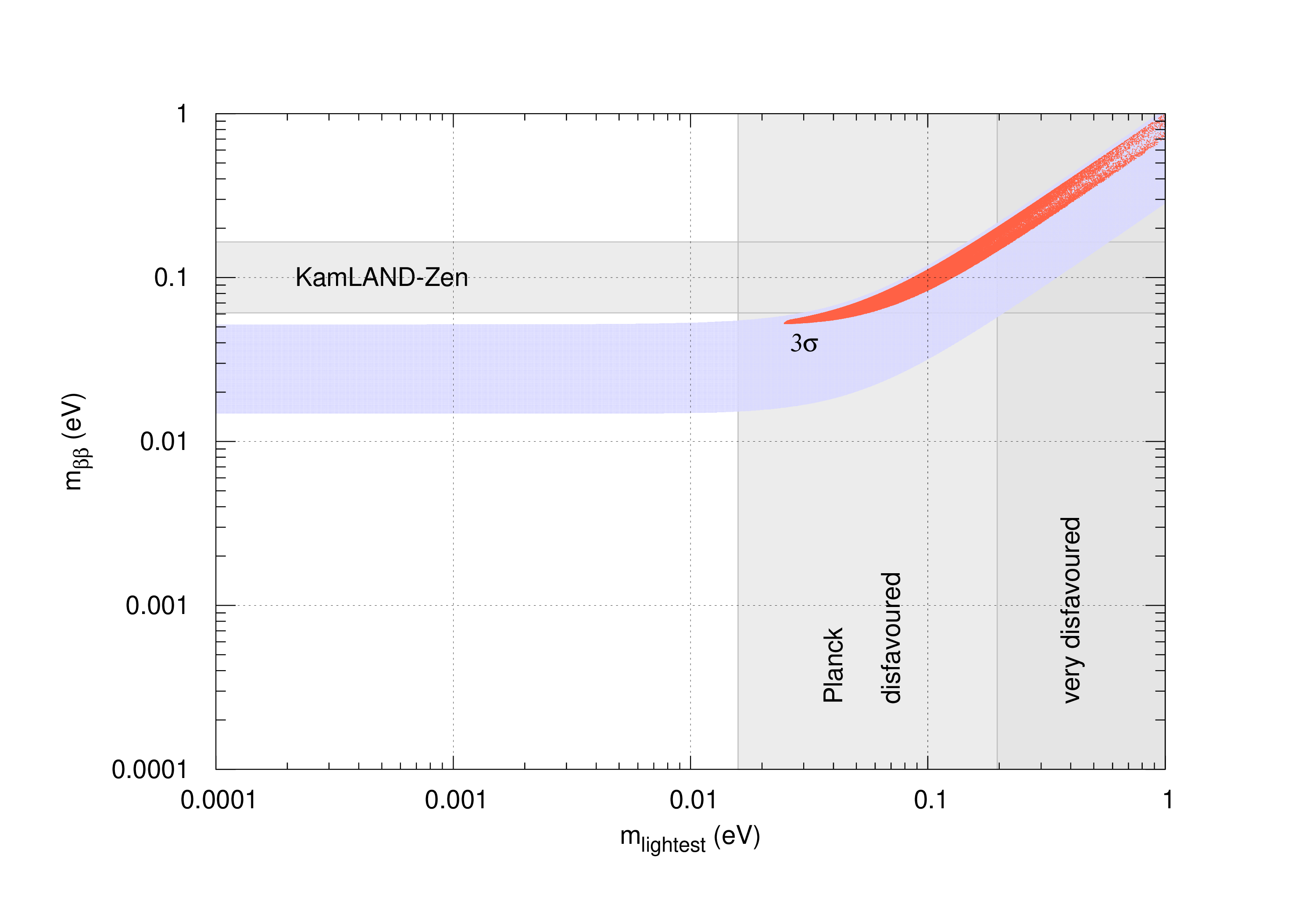}
         \caption{Inverted Ordering}
		 \label{fig:mee weinbergL3p IO A5}
     \end{subfigure}
        \caption{Predictions of $m_{\text{lightest}}$ vs $m_{\beta\beta}$ for both ordering of neutrino masses compatible with $1\sigma$ (dark-red, NO only, except $\theta_{12}$) and $3\sigma$ data from \cite{Esteban:2020cvm}. In both figures there were also included the current upper limit from KamLAND-Zen $m_{\beta\beta} < 61-165$ meV \cite{KamLAND-Zen:2016pfg} and cosmological constraints from PLANCK 2018 (disfavoured region 0.12 eV $ <\sum m_i <$ 0.60 eV and very disfavoured region $\sum m_i > 0.60$ eV) \cite{Aghanim:2018eyx}.
        In both figures there were also included the current upper limit from KamLAND-Zen $m_{\beta\beta} < 61-165$ meV \cite{KamLAND-Zen:2016pfg} and cosmological constraints from PLANCK 2018 (disfavoured region 0.12 eV $ <\sum m_i <$ 0.60 eV and very disfavoured region $\sum m_i > 0.60$ eV) \cite{Aghanim:2018eyx}.}
        \label{fig:mee weinbergL3p A5}
\end{figure}

It is also possible to obtain the expected $m_{\beta\beta}$ for neutrinoless beta decay using the formula
\begin{align}
m_{\beta\beta} & = |(M_{\nu})_{(1,1)}| \nonumber \\
& = \left| \frac{ 5 + \sqrt{5} } {10} m_1 e^{-2 i \alpha_1}  \cos^2\theta + \frac{2 m_2 e^{-2 i \alpha_3} } { 5 + \sqrt{5} } + \frac{ 5 + \sqrt{5} } { 10 } m_3 e^{2 i \alpha_2} \sin^2\theta \right|,
\label{eq:neutrinoless weinbergL3p A5}
\end{align}
where $m_2$ is given by Eq.(\ref{eq:sum rule weinbergL3p A5}).
Doing a numerical computation, the allowed regions of $m_{\text{lightest}}$ vs $m_{\beta\beta}$ of 
FIG.\ref{fig:mee weinbergL3p A5} 
(for NO, $m_{\text{lightest}} = m_1$ and for IO, $m_{\text{lightest}} = m_3$) were obtained, using again as constraints the data from \cite{Esteban:2020cvm}. 
In both figures it is also shown the current upper limit provided by KamLAND-Zen, $m_{\beta\beta} < 61-165$ meV \cite{KamLAND-Zen:2016pfg}. 
Results from PLANCK 2018 also constrain the sum of neutrino masses, although different constrains can be obtained depending on the data considered (for more details, see \cite{Aghanim:2018eyx}). 
In the figures are plotted two shadowed regions, a very disfavoured region $\sum m_i > 0.60$ eV (considering the limit 95\%C.L.,Planck lensing$+$BAO$+\theta_{MC}$) and a disfavoured region $\sum m_i> 0.12$ eV (considering the limit 95\%C.L.,Planck TT,TE,EE$+$lowE$+$lensing$+$BAO$+\theta_{MC}$). 
These constraints on $\sum m_i$ can be expressed as constraints on $m_{\text{lightest}}$ using the best fit value for the squared mass differences: $m_{\text{lightest}}>0.198$ eV and $m_{\text{lightest}}>0.030$ eV for NO and $m_{\text{lightest}}>0.196$ eV and $m_{\text{lightest}}>0.016$ eV for IO, for the very disfavoured and the disfavoured regions respectively.
We conclude then that both fits in 
TABLE \ref{tab:fitresults weinbergL3p A5} are in the disfavoured region.

For NO, the points that were compatible with the $1\sigma$ regions for the observables other than $\theta_{12}$ were plotted with a darker red colour. 
Only for normal mass orderings do we have points outside the disfavoured region.
The minimum values considering the $3\sigma$ ranges are
\begin{align}
(m_{\text{lightest}})_{\text{min}}^{\text{NO}}\approx  0.015\text{ eV}~~&~~(m_{\beta\beta})_{\text{min}}^{\text{NO}}\approx 0.008 \text{ eV}\nonumber
\\
(m_{\text{lightest}})_{\text{min}}^{\text{IO}}\approx 0.025\text{ eV}~~&~~(m_{\beta\beta})_{\text{min}}^{\text{IO}}\approx 0.052 \text{ eV},
\end{align}
and the minimum and maximum values for the points in dark-red are
\begin{align}
(m_{\text{lightest}})_{\text{min}}^{\text{NO}}\approx  0.017\text{ eV}~~&~~(m_{\beta\beta})_{\text{min}}^{\text{NO}}\approx 0.009 \text{ eV} \nonumber
\\
(m_{\text{lightest}})_{\text{max}}^{\text{NO}}\approx 0.085\text{ eV}~~&~~(m_{\beta\beta})_{\text{max}}^{\text{NO}}\approx 0.074 \text{ eV}.
\end{align}
Taking these considerations into account, we conclude that NO is the preferred mass ordering.

\section{Conclusion \label{sec:conclusion}}

In this paper we make use of the framework of multiple modular symmetries and build a model with the viable Golden Ratio 2 mixing. Relying on a minimal field content, we describe how to break the multiple $A_5$ to a single modular symmetry. We present choices of representations and weights for the fields that produce the model, which leads to leptonic mixing matrix that fits well with the observed values of the angles in a predictive manner. We predict neutrinoless double beta decay. Inverted ordering is disfavoured by cosmological observations and is also disfavoured by the fits to the experimental leptonic mixing observables, with the normal ordering leading to a better fit to the experimental observables.

\section*{Acknowledgments}
IdMV acknowledges funding from Funda\c{c}\~{a}o para a Ci\^{e}ncia e a Tecnologia (FCT) through the contract UID/FIS/00777/2020 and was supported in part by FCT through projects CFTP-FCT Unit 777 (UID/FIS/00777/2019), PTDC/FIS-PAR/29436/2017, CERN/FIS-PAR/0004/2019 and CERN/FIS-PAR/0008/2019 which are partially funded through POCTI (FEDER), COMPETE, QREN and EU.
JL acknowledges funding from Funda\c{c}\~{a}o para a Ci\^{e}ncia e a Tecnologia (FCT) through project CERN/FIS-PAR/0008/2019 and NuPhys - CERN/FIS-PAR/0004/2019.

\appendix

\section{\texorpdfstring{$A_5$}{A5} multiplication rules \label{secapp:A5}}

The group $A_5$ is the group of even permutations of five objects and is the symmetry group of the icosahedron and its dual solid the dodecahedron. It has 60 elements and two generators, S and T:
\begin{equation}
S^2=(ST)^3=T^5=1.
\end{equation}

$A_5$ has five conjugacy classes: 
\begin{align}
C_1&=\{e\} &\\
C_2&=\{T^3ST^2ST, ST^2ST^3, ST^2ST^2ST, ST^3ST, T^3ST^3, T^2ST^2, TS, TSTS, ST^3STS, T^2ST^2ST, \nonumber\\
& \hspace{1.5cm} STST^3, T^3ST, ST^3ST^2, T^3ST^2S, T^3STS, TST^3, ST, STST, TST^3ST, ST^2ST^2S\} \\
C_3&=\{STST^2, T^2ST^3STS, ST^3ST^2S, T^2ST^3, S, ST^3ST^2ST, ST^2ST^3ST, T^2ST^3ST^2, \nonumber\\
& \hspace{1.5cm} STST^3ST^2, TST^2S, ST^2ST^3ST^2, ST^2ST, T^3ST^2, T^2STS, TST^3ST^2S\} \\
C_4&=\{T, T^4, TST, STS, STST^2S, TST^2, T^3S, ST^2, T^2S, ST^3, ST^2STS, T^2ST\} \\
C_5&=\{T^2, T^3, ST^2S, TST^2ST, STST^3ST^2S, TST^3ST^2, STST^3ST, \nonumber\\ 
& \hspace{1.5cm} ST^2ST^2, T^2ST^2S, TST^3STS, T^2ST^3ST, ST^2ST^3STS\}
\end{align}
This group has five irreducible representations: an invariant singlet $\bfs$, two triplets $\bft$ and $\bftp$, a quadruplet $\bfqua$ and a quintuplet $\bfqui$. 
The representations for the generators are in 
TABLE \ref{table:representations A5}.

\begin{table}[ht]
\centering
{\renewcommand{\arraystretch}{1.3}
\centering
\begin{tabular}{c|c|c} 
\toprule
& $S$ & $T$ \\
\colrule
$\bfs$ & 1 & 1 \\
$\bft$ & $\frac{1}{\sqrt{5}}
\begin{pmatrix} 
1 & -\sqrt{2} & -\sqrt{2} \\ 
-\sqrt{2} & -\phi & 1/\phi \\ 
-\sqrt{2} & 1/\phi & -\phi 
\end{pmatrix}$ & 
$\begin{pmatrix}
1 & 0 & 0 \\
0 & \zeta & 0 \\
0 & 0 & \zeta^4
\end{pmatrix}$ \\
$\bftp$ & $\frac{1}{\sqrt{5}}
\begin{pmatrix} 
-1 & \sqrt{2} & \sqrt{2} \\ 
\sqrt{2} & -1/\phi & \phi \\ 
\sqrt{2} & \phi & -1/\phi 
\end{pmatrix}$ & 
$\begin{pmatrix} 
1 & 0 & 0 \\ 
0 & \zeta^2 & 0 \\ 
0 & 0 & \zeta^3 
\end{pmatrix}$ \\
$\bfqua$ & $\frac{1}{\sqrt{5}}
\begin{pmatrix} 
1 & 1/\phi & \phi & -1 \\ 
1/\phi & -1 & 1 & \phi \\ 
\phi & 1 & -1 & 1/\phi \\ 
-1 & \phi & 1/\phi & 1
\end{pmatrix}$ & 
$\begin{pmatrix} 
\zeta & 0 & 0 & 0\\ 
0 & \zeta^2 & 0 & 0\\ 
0 & 0 & \zeta^3 & 0 \\
0 & 0 & 0 & \zeta^4 \\
\end{pmatrix}$ \\
$\bfqui$ & $\frac{1}{\sqrt{5}}
\begin{pmatrix} 
-1 & \sqrt{6} & \sqrt{6} & \sqrt{6} & \sqrt{6} \\ 
\sqrt{6} & 1/\phi^2 & -2\phi & 2/\phi & \phi^2 \\ 
\sqrt{6} & -2\phi & \phi^2 & 1/\phi^2 & 2/\phi \\ 
\sqrt{6} & 2/\phi & 1/\phi^2 & \phi^2 & 2\phi \\ 
\sqrt{6} & \phi^2 & 2/\phi & -2\phi & 1/\phi^2
\end{pmatrix}$ & 
$\begin{pmatrix} 
1 & 0 & 0 & 0 & 0 \\
0 & \zeta & 0 & 0 & 0\\ 
0 & 0 & \zeta^2 & 0 & 0\\ 
0 & 0 & 0 & \zeta^3 & 0 \\
0 & 0 & 0 & 0 & \zeta^4 \\
\end{pmatrix}$ \\
\botrule
\end{tabular}}
\caption{Representation for the two generators of $A_5$, where $\phi = \frac{1 + \sqrt{5}}{2}$ and $\zeta = e^{2\pi i/5}$.}
\label{table:representations A5}
\end{table}

The product of two irreps decomposes in the following way: 
\begin{align}
&\bft \otimes \bft = \bfs \oplus  \bft \oplus \bfqui \\
&\bft \otimes \bftp = \bfqua \oplus  \bfqui \\
&\bft \otimes \bfqua = \bftp \oplus  \bfqua \oplus \bfqui \\
&\bft \otimes \bfqui = \bft \oplus  \bftp \oplus \bfqua \oplus \bfqui \\
&\bftp \otimes \bftp = \bfs \oplus  \bftp \oplus \bfqui \\
&\bftp \otimes \bfqua = \bft \oplus  \bfqua \oplus \bfqui \\
&\bftp \otimes \bfqui = \bft \oplus  \bftp \oplus \bfqua \oplus \bfqui \\
&\bfqua \otimes \bfqua = \bfs \oplus \bft \oplus  \bftp \oplus \bfqua \oplus \bfqui \\
&\bfqua \otimes \bfqui = \bft \oplus \bftp \oplus  \bfqua \oplus \bfqui_1 \oplus \bfqui_2 \\
&\bfqui \otimes \bfqui = \bfs \oplus \bft \oplus \bftp \oplus  \bfqua_1 \oplus \bfqua_2 \oplus \bfqui_1 \oplus \bfqui_2
\end{align}
The factors considered for the representation in TABLE \ref{table:representations A5} lead to the following decomposition, with the Clebsch-Gordan coefficients in \cite{Novichkov:2018nkm}:
\begin{align}
\begin{pmatrix}
a_1\\
a_2\\
a_3
\end{pmatrix}_\bft
\otimes
\begin{pmatrix}
b_1\\
b_2\\
b_3
\end{pmatrix}_\bft
& = (a_1b_1+a_2b_3+a_3b_2)_\bfs
\oplus
\begin{pmatrix}
a_2b_3-a_3b_2\\
a_1b_2-a_2b_1\\
a_3b_1-a_1b_3
\end{pmatrix}_\bft
\oplus
\begin{pmatrix}
2a_1b_1-a_2b_3-a_3b_2 \\
-\sqrt{3}a_1b_2-\sqrt{3}a_2b_1 \\
\sqrt{6}a_2b_2 \\
\sqrt{6}a_3b_3 \\
-\sqrt{3}a_1b_3-\sqrt{3}a_3b_1
\end{pmatrix}_\bfqui
\\
\begin{pmatrix}
a_1\\
a_2\\
a_3
\end{pmatrix}_\bftp
\otimes
\begin{pmatrix}
b_1\\
b_2\\
b_3
\end{pmatrix}_\bftp
& = (a_1b_1+a_2b_3+a_3b_2)_\bfs
\oplus
\begin{pmatrix}
a_2b_3-a_3b_2\\
a_1b_2-a_2b_1\\
a_3b_1-a_1b_3
\end{pmatrix}_\bftp
\oplus
\begin{pmatrix}
2a_1b_1-a_2b_3-a_3b_2 \\
\sqrt{6}a_3b_3 \\
-\sqrt{3}a_1b_2-\sqrt{3}a_2b_1 \\
-\sqrt{3}a_1b_3-\sqrt{3}a_3b_1 \\
\sqrt{6}a_2b_2
\end{pmatrix}_\bfqui
\\
\begin{pmatrix}
a_1\\
a_2\\
a_3
\end{pmatrix}_\bft
\otimes
\begin{pmatrix}
b_1\\
b_2\\
b_3
\end{pmatrix}_\bftp
& =
\begin{pmatrix}
\sqrt{2}a_2b_1+a_3b_2\\
-\sqrt{2}a_1b_2-a_3b_3\\
-\sqrt{2}a_1b_3-a_2b_2\\
\sqrt{2}a_3b_1+a_2b_3
\end{pmatrix}_\bfqua
\oplus
\begin{pmatrix}
\sqrt{3}a_1b_1 \\
a_2b_1+\sqrt{2}a_3b_2\\
a_1b_2-\sqrt{2}a_3b_3\\
a_1b_3-\sqrt{2}a_2b_2\\
a_3b_1+\sqrt{2}a_2b_3
\end{pmatrix}_\bfqui
\\
\begin{pmatrix}
a_1\\
a_2\\
a_3
\end{pmatrix}_\bft
\otimes
\begin{pmatrix}
b_1\\
b_2\\
b_3\\
b_4
\end{pmatrix}_\bfqua
& =
\begin{pmatrix}
-\sqrt{2}a_2b_4-\sqrt{2}a_3b_1\\
\sqrt{2}a_1b_2-a_2b_1+a_3b_3\\
\sqrt{2}a_1b_3-a_3b_4+a_2b_2\\
\end{pmatrix}_\bftp
\oplus
\begin{pmatrix}
a_1b_1-\sqrt{2}a_3b_2\\
-a_1b_2-\sqrt{2}a_2b_1\\
a_1b_3+\sqrt{2}a_3b_4\\
-a_1b_4+\sqrt{2}a_2b_3
\end{pmatrix}_\bfqua
\oplus \nonumber \\
&\oplus
\begin{pmatrix}
\sqrt{6}a_2b_4 - \sqrt{6}a_3b_1 \\
2\sqrt{2}a_1b_1+2a_3b_2 \\
-\sqrt{2}a_1b_2+a_2b_1+3a_3b_3 \\
\sqrt{2}a_1b_3-a_3b_4-3a_2b_2 \\
-2\sqrt{2}a_1b_4-2a_2b_3
\end{pmatrix}_\bfqui 
\\
\begin{pmatrix}
a_1\\
a_2\\
a_3
\end{pmatrix}_\bftp
\otimes
\begin{pmatrix}
b_1\\
b_2\\
b_3\\
b_4
\end{pmatrix}_\bfqua
& =
\begin{pmatrix}
-\sqrt{2}a_2b_3-\sqrt{2}a_3b_2\\
\sqrt{2}a_1b_1+a_2b_4-a_3b_3\\
\sqrt{2}a_1b_4+a_3b_1-a_2b_2\\
\end{pmatrix}_\bft
\oplus
\begin{pmatrix}
a_1b_1+\sqrt{2}a_3b_3\\
a_1b_2-\sqrt{2}a_3b_4\\
-a_1b_3+\sqrt{2}a_2b_1\\
-a_1b_4-\sqrt{2}a_2b_2
\end{pmatrix}_\bfqua
\oplus \nonumber \\
&\oplus
\begin{pmatrix}
\sqrt{6}a_2b_3 - \sqrt{6}a_3b_2 \\
\sqrt{2}a_1b_1-3a_2b_4-a_3b_3 \\
2\sqrt{2}a_1b_2+2a_3b_4 \\
-2\sqrt{2}a_1b_3-2a_2b_1 \\
-\sqrt{2}a_1b_4+3a_3b_1+a_2b_2
\end{pmatrix}_\bfqui 
\end{align}
\begin{align}
\begin{pmatrix}
a_1\\
a_2\\
a_3
\end{pmatrix}_\bft
\otimes
\begin{pmatrix}
b_1\\
b_2\\
b_3\\
b_4\\
b_5
\end{pmatrix}_\bfqui
& =
\begin{pmatrix}
-2a_1b_1+\sqrt{3}a_2b_5+\sqrt{3}a_3b_2\\
\sqrt{3}a_1b_2+a_2b_1-\sqrt{6}a_3b_3\\
\sqrt{3}a_1b_5+a_3b_1-\sqrt{6}a_2b_4\\
\end{pmatrix}_\bft
\oplus
\begin{pmatrix}
\sqrt{3}a_1b_1+a_2b_5+a_3b_2\\
a_1b_3-\sqrt{2}a_2b_2-\sqrt{2}a_3b_4\\
a_1b_4-\sqrt{2}a_2b_3-\sqrt{2}a_3b_5\\
\end{pmatrix}_\bftp
\oplus \nonumber \\
& \oplus
\begin{pmatrix}
2\sqrt{2}a_1b_2-\sqrt{6}a_2b_1+a_3b_3 \\
-\sqrt{2}a_1b_3+2a_2b_2-3a_3b_4 \\
\sqrt{2}a_1b_4-2a_2b_5+3a_2b_3 \\
-2\sqrt{2}a_1b_5+\sqrt{6}a_3b_1-a_2b_4
\end{pmatrix}_\bfqua
\oplus
\begin{pmatrix}
\sqrt{3}a_2b_5 - \sqrt{3}a_3b_2 \\
-a_1b_2-\sqrt{3}a_2b_1-\sqrt{2}a_3b_3 \\
-2a_1b_3-\sqrt{2}a_2b_2 \\
2a_1b_4+\sqrt{2}a_3b_5 \\
a_1b_5+\sqrt{3}a_3b_1+\sqrt{2}a_2b_4
\end{pmatrix}_\bfqui 
\\
\begin{pmatrix}
a_1\\
a_2\\
a_3
\end{pmatrix}_\bftp
\otimes
\begin{pmatrix}
b_1\\
b_2\\
b_3\\
b_4\\
b_5
\end{pmatrix}_\bfqui
& =
\begin{pmatrix}
\sqrt{3}a_1b_1+a_2b_4+a_3b_3\\
a_1b_2-\sqrt{2}a_2b_5-\sqrt{2}a_3b_4\\
a_1b_5-\sqrt{2}a_2b_3-\sqrt{2}a_3b_2\\
\end{pmatrix}_\bft
\oplus
\begin{pmatrix}
-2a_1b_1+\sqrt{3}a_2b_4+\sqrt{3}a_3b_3\\
\sqrt{3}a_1b_3+a_2b_1-\sqrt{6}a_3b_5\\
\sqrt{3}a_1b_4+a_3b_1-\sqrt{6}a_2b_2\\
\end{pmatrix}_\bftp
\oplus \nonumber \\
& \oplus
\begin{pmatrix}
\sqrt{2}a_1b_2-2a_3b_4+3a_2b_5 \\
2\sqrt{2}a_1b_3-\sqrt{6}a_2b_1+a_3b_5 \\
-2\sqrt{2}a_1b_4+\sqrt{6}a_3b_1-a_2b_2 \\
-\sqrt{2}a_1b_5+2a_2b_3-3a_3b_2
\end{pmatrix}_\bfqua
\oplus
\begin{pmatrix}
\sqrt{3}a_2b_4 - \sqrt{3}a_3b_3 \\
2a_1b_2+\sqrt{2}a_3b_4 \\
-a_1b_3-\sqrt{3}a_2b_1-\sqrt{2}a_3b_5 \\
a_1b_4+\sqrt{3}a_3b_1+\sqrt{2}a_2b_2 \\
-2a_1b_5-\sqrt{2}a_2b_3
\end{pmatrix}_\bfqui 
\\
\begin{pmatrix}
a_1\\
a_2\\
a_3\\
a_4
\end{pmatrix}_\bfqua
\otimes
\begin{pmatrix}
b_1\\
b_2\\
b_3\\
b_4
\end{pmatrix}_\bfqua
& = (a_1b_4+a_2b_3+a_3b_2+a_4b_1)_\bfs
\oplus
\begin{pmatrix}
a_2b_3-a_3b_2+a_4b_1-a_1b_4\\
\sqrt{2}a_2b_4-\sqrt{2}a_4b_2\\
\sqrt{2}a_1b_3-\sqrt{2}a_3b_1
\end{pmatrix}_\bft
\oplus \nonumber \\
&\oplus
\begin{pmatrix}
a_2b_3-a_3b_2+a_1b_4-a_4b_1\\
\sqrt{2}a_3b_4-\sqrt{2}a_4b_3\\
\sqrt{2}a_1b_2-\sqrt{2}a_2b_1
\end{pmatrix}_\bft
\oplus
\begin{pmatrix}
a_3b_3+a_2b_4+a_4b_2\\
a_1b_1+a_3b_4+a_4b_3\\
a_4b_4+a_1b_2+a_2b_1\\
a_2b_2+a_1b_3+a_3b_1\\
\end{pmatrix}_\bfqua
\oplus \nonumber \\
&\oplus
\begin{pmatrix}
\sqrt{3}a_1b_4+\sqrt{3}a_4b_1-\sqrt{3}a_2b_3-\sqrt{3}a_3b_2 \\
2\sqrt{2}a_3b_3-\sqrt{2}a_2b_4-\sqrt{2}a_4b_2 \\
-2\sqrt{2}a_1b_1+\sqrt{2}a_3b_4+\sqrt{2}a_4b_3 \\
-2\sqrt{2}a_4b_4+\sqrt{2}a_1b_2-\sqrt{2}a_2b_1 \\
2\sqrt{2}a_2b_2-\sqrt{2}a_1b_3-\sqrt{2}a_3b_1
\end{pmatrix}_\bfqui
\end{align}

\begin{align}
\begin{pmatrix}
a_1\\
a_2\\
a_3\\
a_4
\end{pmatrix}_\bfqua
\otimes
\begin{pmatrix}
b_1\\
b_2\\
b_3\\
b_4\\
b_5
\end{pmatrix}_\bfqui
& =
\begin{pmatrix}
2\sqrt{2}a_1b_5-2\sqrt{2}a_4b_2+\sqrt{2}a_3b_3-\sqrt{2}a_2b_4 \\
3a_3b_4+2a_2b_5-a_4b_3-\sqrt{6}a_1b_1 \\
-3a_2b_3-2a_3b_2+a_1b_4+\sqrt{6}a_4b_1
\end{pmatrix}_\bft
\oplus \nonumber \\
&\oplus
\begin{pmatrix}
2\sqrt{2}a_2b_4-2\sqrt{2}a_3b_3+\sqrt{2}a_1b_5-\sqrt{2}a_4b_2 \\
3a_1b_2+2a_4b_4-a_3b_5-\sqrt{6}a_2b_1 \\
-3a_4b_5-2a_1b_3+a_2b_2+\sqrt{6}a_3b_1
\end{pmatrix}_\bftp
\oplus \nonumber \\
&\oplus
\begin{pmatrix}
\sqrt{3}a_1b_1-\sqrt{2}a_2b_5+\sqrt{2}a_3b_4-2\sqrt{2}a_4b_3 \\
-\sqrt{2}a_1b_2-\sqrt{3}a_2b_1+2\sqrt{2}a_3b_5+\sqrt{2}a_4b_4 \\
\sqrt{2}a_1b_3+2\sqrt{2}a_2b_2-\sqrt{3}a_3b_1-\sqrt{2}a_4b_5 \\
-2\sqrt{2}a_1b_4+\sqrt{2}a_2b_3-\sqrt{2}a_3b_2+\sqrt{3}a_4b_1
\end{pmatrix}_\bfqua
\oplus \nonumber \\
&\oplus
\begin{pmatrix}
\sqrt{2}a_1b_5-\sqrt{2}a_2b_4-\sqrt{2}a_3b_3+\sqrt{2}a_4b_2 \\
-\sqrt{2}a_1b_1-\sqrt{3}a_3b_4-\sqrt{3}a_4b_3 \\
\sqrt{3}a_1b_2+\sqrt{2}a_2b_1+\sqrt{3}a_3b_5 \\
\sqrt{3}a_2b_2+\sqrt{2}a_3b_1+\sqrt{3}a_4b_5 \\
-\sqrt{3}a_1b_4-\sqrt{3}a_2b_3-\sqrt{2}a_4b_1
\end{pmatrix}_{\bfqui_1}
\oplus \nonumber \\
&\oplus
\begin{pmatrix}
2a_1b_5+4a_2b_4+4a_3b_3+2a_4b_2 \\
4a_1b_1+2\sqrt{6}a_2b_5 \\
-\sqrt{6}a_1b_2+2a_2b_1-\sqrt{6}a_3b_5+2\sqrt{6}a_4b_4 \\
2\sqrt{6}a_1b_3-\sqrt{6}a_2b_2+2a_3b_1-\sqrt{6}a_4b_5 \\
2\sqrt{6}a_3b_2+4a_4b_1
\end{pmatrix}_{\bfqui_2}
\\
\begin{pmatrix}
a_1\\
a_2\\
a_3\\
a_4\\
a_5
\end{pmatrix}_\bfqui
\otimes
\begin{pmatrix}
b_1\\
b_2\\
b_3\\
b_4\\
b_5
\end{pmatrix}_\bfqui
& = (a_1b_1+a_2b_5+a_5b_2+a_3b_4+a_4b_3)_\bfs
\oplus \nonumber \\
&\oplus
\begin{pmatrix}
a_2b_5-a_5b_2+2a_3b_4-2a_4b_3\\
\sqrt{3}a_2b_1-\sqrt{3}a_1b_2+\sqrt{2}a_3b_5-\sqrt{2}a_5b_3\\
\sqrt{3}a_1b_5-\sqrt{3}a_5b_1+\sqrt{2}a_2b_4-\sqrt{2}a_2b_4
\end{pmatrix}_\bft
\oplus \nonumber \\
&\oplus
\begin{pmatrix}
a_4b_3-a_3b_4+2a_2b_5-2a_5b_2\\
\sqrt{3}a_1b_3-\sqrt{3}a_3b_1+\sqrt{2}a_4b_5-\sqrt{2}a_5b_4\\
\sqrt{3}a_4b_1-\sqrt{3}a_1b_4+\sqrt{2}a_2b_3-\sqrt{2}a_3b_4
\end{pmatrix}_\bftp
\oplus \nonumber \\
&\oplus
\begin{pmatrix}
4\sqrt{3}a_4b_4+3\sqrt{2}a_1b_2+3\sqrt{2}a_2b_1-\sqrt{3}a_3b_5-\sqrt{3}a_5b_3 \\
4\sqrt{3}a_2b_2+3\sqrt{2}a_1b_3+3\sqrt{2}a_3b_1-\sqrt{3}a_4b_5-\sqrt{3}a_5b_4 \\
4\sqrt{3}a_5b_5+3\sqrt{2}a_1b_4+3\sqrt{2}a_4b_1-\sqrt{3}a_3b_2-\sqrt{3}a_2b_3 \\
4\sqrt{3}a_3b_3+3\sqrt{2}a_1b_5+3\sqrt{2}a_5b_1-\sqrt{3}a_2b_4-\sqrt{3}a_4b_2
\end{pmatrix}_{\bfqua_1}
\oplus \nonumber \\
&\oplus
\begin{pmatrix}
\sqrt{2}a_1b_2-\sqrt{2}a_2b_1+\sqrt{3}a_3b_5-\sqrt{3}a_5b_3 \\
\sqrt{2}a_3b_1-\sqrt{2}a_1b_3+\sqrt{3}a_4b_5-\sqrt{3}a_5b_4 \\
\sqrt{2}a_4b_1-\sqrt{2}a_1b_4+\sqrt{3}a_3b_2-\sqrt{3}a_2b_3 \\
\sqrt{2}a_1b_5-\sqrt{2}a_5b_1+\sqrt{3}a_4b_2-\sqrt{3}a_2b_4
\end{pmatrix}_{\bfqua_2}
\oplus \nonumber \\
&\oplus
\begin{pmatrix}
2a_1b_1+a_2b_5+a_5b_2-2a_3b_4-2a_4b_3 \\
a_1b_2+a_2b_1+\sqrt{6}a_3b_5+\sqrt{6}a_5b_3 \\
\sqrt{6}a_2b_2-2a_1b_3-2a_3b_1 \\
\sqrt{6}a_5b_5-2a_1b_4-2a_4b_1 \\
a_1b_5+a_5b_1+\sqrt{6}a_2b_4+\sqrt{6}a_4b_2
\end{pmatrix}_{\bfqui_1}
\oplus \nonumber \\
&\oplus
\begin{pmatrix}
2a_1b_1+a_3b_4+a_4b_3-2a_2b_5-2a_5b_2 \\
\sqrt{6}a_4b_4-2a_1b_2-2a_2b_1 \\
a_1b_3+a_3b_1+\sqrt{6}a_4b_5+\sqrt{6}a_5b_4 \\
a_1b_4+a_4b_1+\sqrt{6}a_2b_3+\sqrt{6}a_3b_2 \\
\sqrt{6}a_3b_3-2a_1b_5-2a_5b_1
\end{pmatrix}_{\bfqui_2}
\end{align}

\section{Modular forms of weight 2 for \texorpdfstring{$A_5$}{A5} \label{secapp:modular forms A5}}

The linear space of modular forms of level $N=5$ and weight 2 has dimension 11. These modular functions are arranged into two triplets $\bft$ and $\bftp$ and a quintuplet $\bfqui$ of $\Gamma_5$.
Modular forms of higher weight can be constructed from polynomials of these eleven modular functions.

The weight 2 modular functions can be expressed as linear combinations of logarithmic derivatives of some functions $\alpha_{i,j}(\tau)$, closed under the action of $A_5$, and these can be in terms of the theta function $\theta_3(z(\tau),t(\tau))$:
\begin{equation}
\theta_3(z,t)= \sum_{k\in\mathbb{Z}} q^{k^2}e^{2\pi i k z} = 1 + 2 \sum_{k\in\mathbb{N}} q^{k^2}\cos(2\pi k z) \,,\, q=e^{\pi i t}
\end{equation}

The seed functions $\alpha_{i,j}(\tau)$ are explicitly:
\begin{equation}
\begin{aligned}[c]
\alpha_{1,-1}(\tau) &\equiv \theta_3\left( \frac{\tau+1}{2}, 5\tau\right)
, \\
\alpha_{1,0}(\tau) &\equiv \theta_3\left( \frac{\tau+9}{10}, \frac{\tau}{5}\right)
, \\
\alpha_{1,1}(\tau) &\equiv \theta_3\left( \frac{\tau}{10}, \frac{\tau+1}{5}\right)
, \\
\alpha_{1,2}(\tau) &\equiv \theta_3\left( \frac{\tau+1}{10}, \frac{\tau+2}{5}\right)
, \\
\alpha_{1,3}(\tau) &\equiv \theta_3\left( \frac{\tau+2}{10}, \frac{\tau+3}{5}\right)
, \\
\alpha_{1,4}(\tau) &\equiv \theta_3\left( \frac{\tau+3}{10}, \frac{\tau+4}{5}\right)
,
\end{aligned}
\qquad
\begin{aligned}[c]
\alpha_{2,-1}(\tau) &\equiv e^{2 \pi i \tau / 5}  
\theta_3\left( \frac{3\tau+1}{2}, 5\tau\right)
, \\
\alpha_{2,0}(\tau) &\equiv \theta_3\left( \frac{\tau+7}{10}, \frac{\tau}{5}\right)
, \\
\alpha_{2,1}(\tau) &\equiv \theta_3\left( \frac{\tau+8}{10}, \frac{\tau+1}{5}\right)
, \\
\alpha_{2,2}(\tau) &\equiv \theta_3\left( \frac{\tau+9}{10}, \frac{\tau+2}{5}\right)
, \\
\alpha_{2,3}(\tau) &\equiv \theta_3\left( \frac{\tau}{10}, \frac{\tau+3}{5}\right)
, \\
\alpha_{2,4}(\tau) &\equiv \theta_3\left( \frac{\tau+1}{10}, \frac{\tau+4}{5}\right)
.
\end{aligned}
\label{eq:alpha ij theta 3}
\end{equation}

The linear combination of the logarithmic derivatives of these functions,
\begin{align}
Y(c_{1,-1},\ldots,c_{1,4};c_{2,-1},\ldots,c_{2,4}|\tau) \equiv \sum_{i,j} c_{i,j}
\frac{{\rm d}}{{\rm d}\tau}\log\alpha_{i,j}(\tau)
,\quad \textrm{with } \sum_{i,j} c_{i,j} = 0,
\end{align}
span the linear space of the modular forms of level $N=5$ and weight 2. These are then divided into the multiplets:
\begin{align}
Y_\bfqui(\tau) = \begin{pmatrix}
Y_{1}(\tau)\\
Y_{2}(\tau)\\
Y_{3}(\tau)\\
Y_{4}(\tau)\\
Y_{5}(\tau)
\end{pmatrix}
& \equiv
\begin{pmatrix}
-\frac{1}{\sqrt{6}}Y\left(-5,1,1,1,1,1;-5,1,1,1,1,1\middle|\tau\right)\\
Y(0,1,\zeta^4,\zeta^3,\zeta^2,\zeta;0,1,\zeta^4,\zeta^3,\zeta^2,\zeta|\tau)\\
Y(0,1,\zeta^3,\zeta,\zeta^4,\zeta^2;0,1,\zeta^3,\zeta,\zeta^4,\zeta^2|\tau)\\
Y(0,1,\zeta^2,\zeta^4,\zeta,\zeta^3;0,1,\zeta^2,\zeta^4,\zeta,\zeta^3|\tau)\\
Y(0,1,\zeta,\zeta^2,\zeta^3,\zeta^4;0,1,\zeta,\zeta^2,\zeta^3,\zeta^4|\tau)
\end{pmatrix},
\label{eq:mf qui} \\
Y_\bft(\tau) = 
\begin{pmatrix}
Y_{6}(\tau)\\
Y_{7}(\tau)\\
Y_{8}(\tau)
\end{pmatrix}
&\equiv
\begin{pmatrix}
\frac{1}{\sqrt{2}}Y\left(-\sqrt{5},-1,-1,-1,-1,-1;\sqrt{5},1,1,1,1,1\middle|\tau\right)\\
Y(0,1,\zeta^4,\zeta^3,\zeta^2,\zeta;0,-1,-\zeta^4,-\zeta^3,-\zeta^2,-\zeta|\tau)\\
Y(0,1,\zeta,\zeta^2,\zeta^3,\zeta^4;0,-1,-\zeta,-\zeta^2,-\zeta^3,-\zeta^4|\tau)
\end{pmatrix},
\label{eq:mf t} \\
Y_\bftp(\tau) = 
\begin{pmatrix}
Y_{9}(\tau)\\
Y_{10}(\tau)\\
Y_{11}(\tau)
\end{pmatrix}
&\equiv
\begin{pmatrix}
\frac{1}{\sqrt{2}}Y\left(\sqrt{5},-1,-1,-1,-1,-1;-\sqrt{5},1,1,1,1,1\middle|\tau\right)\\
Y(0,1,\zeta^3,\zeta,\zeta^4,\zeta^2;0,-1,-\zeta^3,-\zeta,-\zeta^4,-\zeta^2|\tau)\\
Y(0,1,\zeta^2,\zeta^4,\zeta,\zeta^3;0,-1,-\zeta^2,-\zeta^4,-\zeta,-\zeta^3|\tau)
\end{pmatrix},
\label{eq:mf tp}
\end{align}
where $\zeta = e^{2\pi i/5}$.

\section{Vacuum alignments for bi-quintuplet \texorpdfstring{$\Phi$}{Phi} in \texorpdfstring{$A_5$}{A5}}
\label{secapp:vac aligns A5 bi-qui}

In this Appendix we consider how to align the VEV of the bi-quintuplet $\Phi$. Following from \cite{deMedeirosVarzielas:2019cyj} and 
\cite{deMedeirosVarzielas:2021pug} where an alignment was obtained in the context of $S_4$ and $A_4$ respectively, we add two driving fields, with the properties present in TABLE \ref{tab:driving fields bi-qui A5}.

\begin{table}[ht]
\centering
{\renewcommand{\arraystretch}{1.2}
\begin{tabular}{c||c|c|c|c}
\toprule
Fields & $A_5^l$ & $A_5^{\nu}$ & $2k_l$ & $2k_{\nu}$ \\ 
\colrule
$\chi_{l\nu}$ & $\bfqui$ & $\bfqui$ & 0 & 0 \\ 
$\chi_{l}$ & $\bfqui$ & $\bfs$ & 0 & 0 \\
\botrule
\end{tabular}}
\caption{Transformation properties of the fields responsible for the vacuum alignment of the bi-quintuplet $\Phi$.}
\label{tab:driving fields bi-qui A5}
\end{table}

The superpotential responsible for the vacuum alignment that will be minimised with relation to the driving fields is 
\begin{equation}
w = \Phi \Phi \chi_{l\nu} + M\Phi\chi_{l\nu} + \Phi\Phi\chi_l.
\end{equation}
With this field content, we are only interested in contractions of quintuplets to give quintuplets or singlets. 
Minimising this superpotential in order to the driving fields leads us to the constraints:
\begin{align}
&\sum_{j,k=1,\ldots,5}~\sum_{\beta,\gamma=1,\ldots,5} \Big( 
h_{11} P^{(\bfqui\otimes\bfqui)_{\bfqui_1}}_{ijk} P^{(\bfqui\otimes\bfqui)_{\bfqui_1}}_{\alpha\beta\gamma} 
+ h_{21} P^{(\bfqui\otimes\bfqui)_{\bfqui_2}}_{ijk} P^{(\bfqui\otimes\bfqui)_{\bfqui_1}}_{\alpha\beta\gamma} + \nonumber \\
& \hspace{3.5cm} 
+ h_{12} P^{(\bfqui\otimes\bfqui)_{\bfqui_1}}_{ijk} P^{(\bfqui\otimes\bfqui)_{\bfqui_2}}_{\alpha\beta\gamma} 
+ h_{22} P^{(\bfqui\otimes\bfqui)_{\bfqui_2}}_{ijk} P^{(\bfqui\otimes\bfqui)_{\bfqui_2}}_{\alpha\beta\gamma} \Big)  (\Phi)_{j\beta}(\Phi)_{k\gamma} + M(\Phi)_{i\alpha} = 0, \\
&\hspace{1cm}~\text{for}~i=1,\ldots,5, \alpha=1,\ldots,5 \nonumber \\
& \sum_{j,k=1,\ldots,5}~\sum_{\alpha,\beta=1,\ldots,5}P^{(\bfqui\otimes\bfqui)_\bfs}_{\alpha\beta} \left( h_{1} P^{(\bfqui\otimes\bfqui)_{\bfqui_1}}_{ijk} + h_{2} P^{(\bfqui\otimes\bfqui)_{\bfqui_2}}_{ijk} \right) (\Phi)_{j\alpha}(\Phi)_{k\beta} = 0 ~~\text{for}~~ i=1,\ldots,5.
\label{eq:vac align A5 bi-qui constraints}
\end{align}
where $5 \times 5$ matrices that describe the multiplication rules in Section \ref{secapp:A5} were introduced:
\begin{align}
P^{(\bfqui\otimes\bfqui)_\bfs} & = 
\begin{pmatrix}
1 & 0 & 0 & 0 & 0 \\
0 & 0 & 0 & 0 & 1 \\
0 & 0 & 0 & 1 & 0 \\
0 & 0 & 1 & 0 & 0 \\
0 & 1 & 0 & 0 & 0 \\
\end{pmatrix}\\
P^{(\bfqui\otimes\bfqui)_{\bfqui_1}} & = \left(
\begin{pmatrix}
2 & 0 & 0 & 0 & 0 \\
0 & 0 & 0 & 0 & 1 \\
0 & 0 & 0 & -2 & 0 \\
0 & 0 & -2 & 0 & 0 \\
0 & 1 & 0 & 0 & 0 \\
\end{pmatrix},
\begin{pmatrix}
0 & 1 & 0 & 0 & 0 \\
1 & 0 & 0 & 0 & 0 \\
0 & 0 & 0 & 0 & \sqrt{6} \\
0 & 0 & 0 & 0 & 0 \\
0 & 0 & \sqrt{6} & 0 & 0 \\
\end{pmatrix},
\begin{pmatrix}
0 & 0 & -2 & 0 & 0 \\
0 & \sqrt{6} & 0 & 0 & 0 \\
-2 & 0 & 0 & 0 & 0 \\
0 & 0 & 0 & 0 & 0 \\
0 & 0 & 0 & 0 & 0 \\
\end{pmatrix}, \right. \nonumber \\ 
& \hspace{2cm} \left.
\begin{pmatrix}
0 & 0 & 0 & -2 & 0 \\
0 & 0 & 0 & 0 & 0 \\
0 & 0 & 0 & 0 & 0 \\
-2 & 0 & 0 & 0 & 0 \\
0 & 0 & 0 & 0 & \sqrt{6} \\
\end{pmatrix},
\begin{pmatrix}
0 & 0 & 0 & 0 & 1 \\
0 & 0 & 0 & \sqrt{6} & 0 \\
0 & 0 & 0 & 0 & 0 \\
0 & \sqrt{6} & 0 & 0 & 0 \\
1 & 0 & 0 & 0 & 0 \\
\end{pmatrix}
\right)\\
P^{(\bfqui\otimes\bfqui)_{\bfqui_2}} & = \left(
\begin{pmatrix}
2 & 0 & 0 & 0 & 0 \\
0 & 0 & 0 & 0 & -2 \\
0 & 0 & 0 & 1 & 0 \\
0 & 0 & 1 & 0 & 0 \\
0 & -2 & 0 & 0 & 0 \\
\end{pmatrix},
\begin{pmatrix}
0 & -2 & 0 & 0 & 0 \\
-2 & 0 & 0 & 0 & 0 \\
0 & 0 & 0 & 0 & 0 \\
0 & 0 & 0 & \sqrt{6} & 0 \\
0 & 0 & 0 & 0 & 0 \\
\end{pmatrix},
\begin{pmatrix}
0 & 0 & 1 & 0 & 0 \\
0 & 0 & 0 & 0 & 0 \\
1 & 0 & 0 & 0 & 0 \\
0 & 0 & 0 & 0 & \sqrt{6} \\
0 & 0 & 0 & \sqrt{6} & 0 \\
\end{pmatrix}, \right. \nonumber \\ 
& \hspace{2cm} \left.
\begin{pmatrix}
0 & 0 & 0 & 1 & 0 \\
0 & 0 & \sqrt{6} & 0 & 0 \\
0 & \sqrt{6} & 0 & 0 & 0 \\
1 & 0 & 0 & 0 & 0 \\
0 & 0 & 0 & 0 & 0 \\
\end{pmatrix},
\begin{pmatrix}
0 & 0 & 0 & 0 & -2 \\
0 & 0 & 0 & 0 & 0 \\
0 & 0 & \sqrt{6} & 0 & 0 \\
0 & 0 & 0 & 0 & 0 \\
-2 & 0 & 0 & 0 & 0 \\
\end{pmatrix}
\right)
\end{align}

The solutions of the constraints in Eq.(\ref{eq:vac align A5 bi-qui constraints}) are the 60 elements of $A_5$ in the five dimensional representation, including the VEV that is used in the main text for $\Phi$: 
\begin{align}
\langle\Phi\rangle=v_{\Phi}~\begin{pmatrix}
1 & 0 & 0 & 0 & 0 \\
0 & 0 & 0 & 0 & 1 \\
0 & 0 & 0 & 1 & 0 \\
0 & 0 & 1 & 0 & 0 \\
0 & 1 & 0 & 0 & 0 \\
\end{pmatrix} = 
v_{\Phi} P_{\pi}.
\label{vev phi solutions A5 bi-qui}
\end{align}

\end{document}